\documentclass[epj,nopacs]{svjour}
\usepackage{epsfig,array}
\usepackage{multirow,color}
\usepackage{epsf,graphicx}
\usepackage{times,latexsym,euscript,amstext,amssymb,amsbsy,amsmath,amsfonts}
\usepackage{mdwlist}\usepackage{epsfig}
\usepackage{slashbox,multirow}
\usepackage[section] {placeins}
\newcommand{\be}{\begin{eqnarray}}
\newcommand{\ee}{\end{eqnarray}}
\newcommand{\bc}{\begin{center}}
\newcommand{\ec}{\end{center}}
\newcommand{\bea}{\begin{eqnarray}}
\newcommand{\eea}{\end{eqnarray}}

\newcommand{\er}{$\pm$}

\newcommand{\beq}{\begin{equation}}
\newcommand{\eeq}{\end{equation}}

\newcommand{\threestar}{\raisebox{0.5ex}{$\star\star\star$}}
\newcommand{\twostar}{\raisebox{0.5ex}{$\star\star$}}
\newcommand{\rs}{\raisebox{0.5ex}{$\star$}}
\newcommand{\rts}{\raisebox{0.6ex}{$\star\star$}}
\newcommand{\rmstar}{\raisebox{0.5ex}{(}*\raisebox{0.5ex}{)}}
\newcommand{\rmtstar}{\raisebox{0.5ex}{(}**\raisebox{0.5ex}{)}}

\def\fun#1#2{\lower3.6pt\vbox{\baselineskip0pt\lineskip.9pt
\ialign{$\mathsurround=0pt#1\hfil##\hfil$\crcr#2\crcr\sim\crcr}}}

\begin{document}

\title{\boldmath Properties of baryon resonances from a multichannel partial wave analysis }
\titlerunning{Properties of baryon resonances from a multichannel partial wave analysis  }
\author{
A.V.~Anisovich$\,^{1,2}$, R. Beck$\,^1$, E.~Klempt$\,^1$,
V.A.~Nikonov$\,^{1,2}$, A.V.~Sarantsev$\,^{1,2}$,  and
U.~Thoma$\,^{1}$ }
\authorrunning{A.V.~Anisovich \it et al.}
\institute{$^1\,$Helmholtz-Institut f\"ur Strahlen- und Kernphysik,
Universit\"at Bonn, Germany\\
$^2\,$Petersburg Nuclear Physics Institute, Gatchina, Russia}

\date{Received: \today / Revised version:}

\abstract{Properties of nucleon and $\Delta$ resonances are derived
from a multichannel partial wave analysis. The statistical
significance of pion and photo-induced inelastic reactions off
protons are studied in a multichannel partial-wave analysis.
 \vspace{1mm}   }

\maketitle

\section{Introduction}
Existence and properties of most $N$ and $\Delta$ resonances listed
in the Review of Particle Properties \cite{Nakamura:2010zzi} were
derived from partial wave analyses  of $\pi N$ elastic and charge
exchange scattering data \cite{Hohler:1979yr,Hohler:1993xq,%
Cutkosky:1980rh,Arndt:2006bf}. Additional information on their decay
modes was obtained from inelastic reactions, from $\pi N\to N\eta,
\Lambda K, \Sigma K$ and from an isobar model study of $\pi N\to
N\pi\pi$; photoproduction experiments provided information on the
photo-coupling. The most recent analysis \cite{Arndt:2006bf} --
based on a larger data set and on very precise data from meson
factories -- found no evidence for the existence of 16 of the 32 $N$
and $\Delta$ resonances below 2.2\,GeV listed in the Baryon Particle
Tables. Obviously, the existing data base was not sufficient to
extract a reliable spectrum of $N$ and $\Delta$ resonances from pion
induced reactions alone.

In the last years, an impressive amount of photo-induced reactions
has been studied at ELSA, GRAAL, Jlab, MAMI, and SPring-8, and the
situation has changed significantly. High-statistics data are
available not only on differential cross sections but also on many
polarization observables. In particular, reactions like $\gamma p\to
p\pi^0$, $n\pi^+$, $p\eta$, $p\pi^0\pi^0$, $p\pi^+\pi^-$,
$p\pi^0\eta$, $\Lambda K^+$, $\Sigma^0 K^+$, and $\Sigma^+K^0_s$
have been studied, some of them in great detail.

In this paper, we give a brief account of the results of the
Bonn-Gatchina (BnGa) multichannel partial wave analysis. Main
results have been reported before
\cite{Anisovich:2009zy,Anisovich:2010an,Anisovich:2011ka,%
Anisovich:2011ye}. We found two classes of solutions, called
BnGa2011-01 and BnGa2011-02, which differ in the number and
properties of some positive-parity nucleon resonances at masses
above 1.9\,GeV. The emphasis of the papers
\cite{Anisovich:2010an,Anisovich:2011ye} was on a discussion of the
alternative solutions, on the new resonances found in the analysis,
and on their physics interpretation. In \cite{Anisovich:2009zy},
amplitudes for pion photoproduction off protons were presented, and
in \cite{Anisovich:2011ka}, the focus was to explore possible
interpretations of a narrow structure in the $N\eta$ mass
distribution. The emphasis here is to provide complete information
on resonances, their masses and widths, their helicity amplitudes,
and their decay properties. The statistical evidence (in terms of
$\chi^2$) is given for each resonance and for their decay modes into
$N\gamma, N\pi, N\eta, \Lambda K$ and $\Sigma K$. The results are
not derived from new fits. Only the error analysis has been improved
by storing several acceptable solutions and by calculating (instead
of estimating) properties and errors from the distribution of all
quantities. Hence the results supersede those of
\cite{Anisovich:2010an,Anisovich:2011ye}.

\section{Data used in the partial wave analysis}
\begin{table}[pb]
\caption{\label{piN_data_table}Fit to the real and imaginary part of elastic $\pi N$ amplitudes and
$\chi^2$ contributions for the solution BG2011-02. The elastic scattering data are fitted jointly
with a larger number of further data in a coupled channel approach. }
\bc\begin{tabular}{ccccc}
\hline\hline\\[-2ex]
$\pi N \rightarrow \pi N$& Wave & $N_{\rm data}$ &$w_i$ &$\chi^2_i/N_{\rm data}$\\[1ex]\hline\\[-2ex]
\cite{Arndt:2006bf}& $S_{11}$ & 112 & 30 & 2.11 \\
& $S_{31}$ & 112 & 20 & 2.19 \\
& $P_{11}$ & 112 & 70 & 1.70 \\
& $P_{31}$ & 104 & 20 & 3.74\\
& $P_{13}$ & 112 & 25 & 1.39 \\
& $P_{33}$ & 120 & 15 & 2.77\\
& $D_{13}$ & 108 & 10 & 2.21\\
& $D_{33}$ & 108 & 12 &3.08 \\
& $D_{15}$ & 104 & 20 & 2.29\\
& $F_{15}$ & 88  & 30 & 1.87\\
& $F_{35}$ & 62 & 20 & 1.64\\
& $F_{37}$ & 72 & 10 & 2.76\\
& $F_{17}$ & 82 & 30 & 1.99\\
& $G_{17}$ & 102& 15 & 2.31\\
[0.5ex]\hline \hline\\[-2ex]
\end{tabular}\ec
\end{table}

\begin{table}[pt] \caption{\label{piN_other_data}Pion
induced reactions fitted in the coupled-channel analysis and $\chi^2$ contributions for the
solution BG2011-02.}
\bc\begin{tabular}{ccccc}
\hline\hline\\[-1ex]
$\pi^- p \rightarrow \eta n$ & Observ. & $N_{\rm data}$&$w_i$ & $\chi^2_i/N_{\rm data}$\\[1ex]\hline\\[-2.3ex]
\cite{Richards:1970cy}&  $d\sigma/d\Omega$ & 70 & 20 &1.47 \\
\cite{Prakhov:2005qb}&  $d\sigma/d\Omega$ & 84 & 30 & 2.98 \\[0.5ex]\hline\\[-2.3ex]
\hline\\
[-2.0ex] $\pi^- p \rightarrow K^0\Lambda$ & Observ. & $N_{\rm data}$&$w_i$ & $\chi^2_i/N_{\rm data}$\\[1ex]\hline\\[-2.3ex]
\cite{Knasel:1975rr}&  $d\sigma/d\Omega$ & 300 & 30 &0.90 \\
\cite{Baker:1978qm,Saxon:1979xu}&  $d\sigma/d\Omega$ & 298 & 30 & 2.30 \\
\cite{Baker:1978qm,Saxon:1979xu}&  $P$ & 355 & 30 & 1.77 \\
\cite{Bell:1983dm}&  $\beta$ & 72 & 70 & 1.06 \\
[0.5ex]\hline\\[-2.3ex]
\hline \\[-2.0ex]
$\pi^+ p \rightarrow K^+\Sigma^+$ & Observ. & $N_{\rm data}$&$w_i$ & $\chi^2_i/N_{\rm data}$\\[1ex]\hline\\[-2.3ex]
\cite{Candlin:1982yv,Crawford:1962zz,Winik:1977mm,Baltay:1961,Carayannopoulos:1965}&  $d\sigma/d\Omega$ & 728 & 35 &1.46 \\
\cite{Candlin:1982yv,Crawford:1962zz,Winik:1977mm,Baltay:1961,Carayannopoulos:1965,Bellamy:1972fa}&  $P$ & 351 & 30 & 1.57 \\
\cite{Candlin:1988pn}&  $\beta$ & 7 &600 & 2.04 \\
[0.5ex]\hline\\[-2.3ex]
\hline\\[-2.0ex]
$\pi^- p \rightarrow K^0\Sigma^0$ & Observ. & $N_{\rm data}$&$w_i$ & $\chi^2_i/N_{\rm data}$\\[1ex]\hline\\[-2.3ex]
\cite{Hart:1979jx}&  $d\sigma/d\Omega$ & 259 & 30 &0.98 \\
\cite{Hart:1979jx}&  $P$ & 95 & 30 & 1.30 \\
\hline \hline\\[-2.3ex]
\end{tabular}\ec
%\end{table}
%\begin{table}[pt]
\caption{\label{3BodyReactions}Reactions leading to 3-body final
states included in the event-based likelihood fits; likelihood
values for the solution BG2011-02. CB stands for CB-ELSA; CBT for
CBELSA/TAPS.\vspace{-4mm}}
\bc\begin{tabular}{lcccc}
\hline\hline\\[-1ex]
\multicolumn{2}{c}{$d\sigma/d\Omega(\pi^-p \rightarrow \pi^0\pi^0
n)$} &\hspace{-3mm} $N_{\rm data}$ &$w_i$& $-\ln L$\\[1ex]\hline\\[-2ex]
T=373 MeV   &&\hspace{-3mm} 5248 & 10& -924\\
\multicolumn{2}{l}{T=472 MeV \hspace{16mm} Crystal }  &\hspace{-3mm} 10641& 5& -2603\\
\multicolumn{2}{l}{T=551 MeV \hspace{15mm}  Ball \cite{Prakhov:2004zv}} &\hspace{-3mm} 41172 & 2.5& -7319\\
\multicolumn{2}{l}{T=655 MeV \hspace{16mm}  (BNL)}  &\hspace{-3mm} 63514 & 2& -15165\\
T=691 MeV & &\hspace{-3mm} 30030 & 3.5& -8156\\
T=748 MeV   &&\hspace{-3mm} 30379 & 4& -6881\\[0.4ex]\hline\\[-2.1ex]
\hspace{-2mm}$d\sigma/d\Omega(\gamma p \rightarrow \pi^0\pi^0 p)$
 &\hspace{-5mm}CB\hspace{2mm}\cite{Thoma:2007bm,Sarantsev:2007bk}&\hspace{-3mm} 110601 & 4 & -26953\\
\hspace{-2mm}$d\sigma/d\Omega(\gamma p \rightarrow \pi^0\pi^0 p)$
 &\hspace{-5mm}CB\hspace{2mm}\cite{Fuchs:2012}&\hspace{-3mm} 10000 & 7 & -5276\\
\hspace{-2mm}$d\sigma/d\Omega(\gamma p \rightarrow \pi^0\eta p)$
&\hspace{-4mm}CB \cite{Horn:2007pp,Weinheimer:2003ng,Horn:2008qv}&
\hspace{-3mm}17468 & 8 &
-5701\\
\hline\hline\\[-2ex]
\multicolumn{2}{c}{} &\hspace{-3mm} $N_{\rm data}$ &$w_i$& $\chi^2/N_{\rm data}$\\[1ex]\hline\\[-2ex]
$\Sigma(\gamma p \rightarrow \pi^0\pi^0 p)$
&\hspace{-5mm}GRAAL\hspace{2mm}\cite{Assafiri_03}
&\hspace{-3mm} 128 & 35 & 1.11\\
$\Sigma(\gamma p \rightarrow \pi^0\eta p)$
&\hspace{-5mm}CBT\hspace{2mm}\cite{Gutz:2008zz}&\hspace{-3mm} 180
& 15 & 2.40\\
 $E(\gamma p \rightarrow \pi^0\pi^0 p)$
&\hspace{-5mm}GDH/A2\hspace{2mm}\cite{Ahrens_07} &\hspace{-3mm} 16 & 35 & 1.26
\\
 $I_c$,$I_s(\gamma p \rightarrow \pi^0\pi^0 p)$
&\hspace{-5mm}CBT\hspace{2mm}\cite{Gutz:2009zh} &\hspace{-3mm} 210 &
10 & 1.45\\
 $I_c$,$I_s(\gamma p \rightarrow \pi^0\pi^0 p)$
&\hspace{-5mm}CBT\hspace{2mm}\cite{Sokhoyan:2011kc} &\hspace{-3mm}
1000 &10 & 1.71\\
\hline\hline
\end{tabular}\ec
%\end{table}
%\begin{table}[pt]
\caption{\label{chisquare-eta}Observables from $\eta$
photoproduction fitted in the coupled-channel analysis and $\chi^2$
contributions for the solution BG2011-02.\vspace{-3mm}}
\bc\begin{tabular}{lcccc}
\hline\hline\\[-1ex]
$\gamma p \rightarrow \eta p$ & Observ. & $N_{\rm data}$&$w_i$ & $\chi^2_i/N_{\rm data}$\\[1ex]\hline\\[-2ex]
\cite{McNicoll:2010qk} Crystal Ball @ MAMI & $d\sigma/d\Omega$ &2400 & 2 &1.30 \\
\cite{Crede:2009zzb} CBT& $d\sigma/d\Omega$ &680 & 40 &1.39 \\
\cite{Bartholomy:2007zz} CB& $d\sigma/d\Omega$ &631 & 20 &1.74 \\
\cite{Ajaka:1998zi} GRAAL& $\Sigma$ &51 & 10 &1.81 \\
\cite{Bartalini:2007fg} GRAAL&  $\Sigma$ &150& 15 &1.19\\
\cite{Elsner:2007hm} CBT&$\Sigma$ &34&20&0.82\\
\hline\\[-2ex]
\hline\\[-2ex]
\end{tabular}\vspace{2mm}\ec
\caption{\label{chisquare}Observables from $\pi$ photoproduction
fitted in the coupled-channel analysis and $\chi^2$ contributions
for the solution BG2011-02.\vspace{-2mm}}
\end{table}

\begin{table}[ph]\vspace{1.5mm}
\bc\begin{tabular}{lcccc}
\hline\hline\\[-1ex]
$\gamma p \rightarrow \pi^0 p$ &\hspace{-3mm} Observ. & $N_{\rm data}$&$w_i$ & $\chi^2_i/N_{\rm data}$\\[1ex]\hline\\[-2ex]
\cite{Fuchs:1996ja} (TAPS@MAMI)&\hspace{-3mm} $d\sigma/d\Omega$  & 1692 & 0.8&1.61 \\
\cite{Ahrens:2002gu,Ahrens:2004pf} (GDH A2)&\hspace{-3mm} $d\sigma/d\Omega$  & 164 & 7&1.19 \\
\cite{Bartalini:2005wx} (GRAAL)&\hspace{-3mm} $d\sigma/d\Omega$  & 861 & 2&1.56 \\
\cite{Bartholomy:2004uz,vanPee:2007tw} (CB)&\hspace{-3mm} $d\sigma/d\Omega$  & 1106 & 3.5&1.59 \\
\cite{Dugger:2007bt} (CLAS)&\hspace{-3mm} $d\sigma/d\Omega$ & 592 & 6 &1.19 \\
\cite{Crede:2011dc} (CBT)&\hspace{-3mm} $d\sigma/d\Omega$ & 540 & 6 &2.01 \\
\cite{Bartalini:2005wx,Barbiellini:1970qu,Gorbenko:1974sz,Gorbenko:1978re,Belyaev:1983xf,%
Blanpied:1992nn,Beck:1997ew,Adamian:2000yi,Blanpied:2001ae}&  $\Sigma$ &1492& 3 &2.65\\
\cite{Sparks:2010vb} (CBT) &  $\Sigma$& 374 & 30 & 1.04\\
\cite{Gorbenko:1974sz,Gorbenko:1978re,Belyaev:1983xf,Booth:1976es,Feller:1976ta,%
Gorbenko:1977rd,Herr:1977vx,Fukushima:1977xj,Bussey:1979wt,Agababian:1989kd,%
Asaturian:1986bj,Bock:1998rk,Maloy:1961qy}&  $T$&389& 8 &3.24\\
\cite{Gorbenko:1974sz,%
Gorbenko:1978re,Belyaev:1983xf,Maloy:1961qy,Gorbenko:1975pz,Kato:1979br,Bratashevsky:1980dk,%
Bratashevsky:1986xz}&  $P$&607& 3 &3.14\\
\cite{Bussey:1979wr,Ahrens:2005zq} &  $G$&75& 5 &1.49\\
\cite{Bussey:1979wr} &  $H$&71& 5 &1.22\\
\cite{Ahrens:2002gu,Ahrens:2004pf} &  $E$&140& 7 &1.03\\
\cite{Bratashevsky:1980dk,Avakyan:1991pj}&  $O_{x'}$&7& 10 &1.14\\
\cite{Bratashevsky:1980dk,Avakyan:1991pj}&  $O_{z'}$&7& 10 &0.35\\\hline\\[-2.3ex]
\hline\\[-2ex]
$\gamma p \rightarrow \pi^+ n$ & Observ. & $N_{\rm data}$&$w_i$ & $\chi^2_i/N_{\rm data}$\\[1ex]\hline\\[-2ex]
\cite{Ecklund:1967zz,Betourne:1968bd,Bouquet:1971cv,Fujii:1971qe,%
Ekstrand:1972rt,Fujii:1976jg,Arai:1977kb,Durwen:1980mq,Althoff:1983te,%
Heise:1988ag,Buechler:1994jg,Dannhausen:2001yz}&  $d\sigma/d\Omega$ & 1583 & 2 &1.33  \\
\cite{Ahrens:2004pf,Ahrens:2006gp} (GDH A2)&  $d\sigma/d\Omega$ & 408 & 14 &0.69  \\
\cite{Dugger:2009pn} (CLAS)&  $d\sigma/d\Omega$ & 484 & 4 &1.12  \\
\cite{Blanpied:2001ae,Taylor:1960dn,Smith:1963zza,Alspector:1972pw,Knies:1974zx,%
Ganenko:1976rf,Bussey:1979ju,Getman:1981qt,Hampe:1980jb,Beck:1999ge,%
Ajaka:2000rj,Bocquet:2001ny}&  $\Sigma$ &899 & 3 &3.46\\
\cite{Bussey:1979ju,Getman:1981qt,Althoff:1973kb,Arai:1973xs,Feller:1974qf,Althoff:1975kt,Genzel:1975tx,%
Althoff:1976gq,Althoff:1977ef,Fukushima:1977xh,Getman:1980pw,%
Fujii:1981kx,Dutz:1996uc}&  $T$&661 & 3 &3.09\\
\cite{Bussey:1979ju,Getman:1981qt,Egawa:1981uj}&  $P$&252 & 3 &2.20\\
\cite{Ahrens:2005zq,Bussey:1980fb,Belyaev:1985sp} &  $G$&86 & 8 &5.47\\
\cite{Bussey:1980fb,Belyaev:1985sp,Belyaev:1986va} &  $H$&128& 3& 3.75\\
\cite{Ahrens:2004pf,Ahrens:2006gp} &  $E$&231& 14 & 1.52\\\hline\\[-2.3ex]
\hline
\end{tabular}
\ec
%\end{table}
%\begin{table}[pt]
\caption{\label{chisquare1}Hyperon photoproduction observables
fitted in the coupled-channel analysis and $\chi^2$ contributions
for the solution BG2011-02.}
\bc\begin{tabular}{lcccc}
\hline\hline\\[-2ex]
$\gamma p \rightarrow K^+ \Lambda$ & Observ. & $N_{\rm data}$&$w_i$ & $\chi^2_i/N_{\rm data}$\\[1ex]\hline\\[-2ex]
\cite{McCracken:2009ra} CLAS&  $d\sigma/d\Omega$&1320 &16 &0.69 \\
\cite{Zegers:2003ux} LEPS&  $\Sigma$ &45& 10 & 2.11\\
\cite{Lleres:2007tx} GRAAL&  $\Sigma$ &66& 8 & 2.95\\
\cite{McCracken:2009ra} CLAS&  $P$&1270& 8   &1.82\\
\cite{Lleres:2007tx} GRAAL&  $P$&66 &10 &0.59\\
\cite{Lleres:2008em} GRAAL&  $T$&66 & 15 &1.62\\
\cite{Bradford:2006ba} CLAS&  $C_x$&160 &15 &1.52\\
\cite{Bradford:2006ba} CLAS&  $C_z$&160 & 15 &1.58\\
\cite{Lleres:2008em} GRAAL&  $O_{x'}$&66 & 12 &1.95\\
\cite{Lleres:2008em} GRAAL&  $O_{z'}$&66 & 15 &1.66\\\hline\\[-2.3ex]
\hline\\[-2ex]
$\gamma p \rightarrow K^+ \Sigma$ & Observ. & $N_{\rm data}$&$w_i$ & $\chi^2_i/N_{\rm data}$\\[1ex]\hline\\[-2ex]
\cite{Dey:2010hh} CLAS& $d\sigma/d\Omega$ & 1590& 3 &1.44 \\
\cite{Zegers:2003ux} LEPS&  $\Sigma$ &45& 10 &1.23\\
\cite{Lleres:2007tx} GRAAL&  $\Sigma$ &42& 10 &1.99\\
\cite{Dey:2010hh} CLAS&  $P$&344 & 12 &2.69\\
\cite{Bradford:2006ba} CLAS&  $C_x$&94 &15 &1.95\\
\cite{Bradford:2006ba} CLAS&  $C_z$&94 &15 &1.66\\\hline\\[-2.3ex]
\hline\\[-2ex]
$\gamma p \rightarrow K^0 \Sigma^+$ & Obsv. & $N_{\rm data}$&$w_i$ & $\chi^2_i/N_{\rm data}$\\[1ex]\hline\\[-2ex]
\cite{Carnahan:2003mk} CLAS&  $d\sigma/d\Omega$ & 48 &3   &3.84 \\
\cite{Lawall:2005np} SAPHIR&  $d\sigma/d\Omega$ & 160 &5 &1.91 \\
\cite{Castelijns:2007qt} CBT&  $d\sigma/d\Omega$ & 72 &10&0.76 \\
\cite{872854} CBT&  $d\sigma/d\Omega$ & 72 &40&0.62 \\
\cite{Castelijns:2007qt} CBT&  $P$&72 & 15 &0.90\\
\cite{872854} CBT&  $P$ & 24 &30&0.94 \\
\cite{872854} CBT&  $\Sigma$ & 15 &50&1.73 \\
\hline\hline
\end{tabular}\ec
\end{table}
Tables~\ref{piN_data_table}-\ref{chisquare1} give an updated list of
the pion- and photo-induced reactions used in the coupled channel
analysis presented here. The data comprise nearly all important
reactions including multiparticle final states. Resonances with
sizable coupling constants to $\pi N$ and $\gamma N$ are thus
unlikely to escape the fits even though further single and double
polarization experiments are certainly needed to unambiguously
constrain the contributing
 amplitudes. A few data sets were omitted
for reasons discussed in \cite{Anisovich:2010an}. The Tables list
the reaction, the observables and references to the data, the number
of data points, the weight with which the data are used in the fits,
and the $\chi^2$ per data point of our final solution BG2011-02.
Multibody final states are fitted in an event-based likelihood fit.
For these reactions, the log likelihood is given (see Eq.
\ref{likeli}). The analysis was constrained by the total cross
sections for $\pi^- p\to n\pi^+\pi^-$ and $\pi^+ p\to p\pi^0\pi^0$
from \cite{Manley:1984jz}.

\section{Partial wave analysis and definitions} The partial wave analysis
method used in this analysis is described in detail in
\cite{Anisovich:2004zz,Anisovich:2006bc}. A shorter survey can be
found in \cite{Anisovich:2010an}. In the Tables below we give pole
parameters as well as Breit-Wigner parameters. Here, we give the
precise definitions used to calculate the quantities given in the
Tables.

The transition amplitude for a pion- or photo-produced reaction from
the initial state $a=\pi N$ or $\gamma N$ and with $b$, e.g.
$\Lambda K^+$, as final state can be defined as
\be
\label{kmatrix}
A_{ab}=K_{ac}(I-i\rho K)_{cb}^{-1}
\ee
where $K$ is called $K$ matrix and $\rho$ is the phase space. A
single resonance is described by a term
\be
K_{ab}=\frac{g_a g_b}{M^2-s}.
\ee
with $g_a, g_b$ being coupling constants.
In this case the equation (\ref{kmatrix}) corresponds to the
relativistic Breit-Wigner amplitude
\be
A_{ab}=\frac{g_a g_b}{M^2-s-i\sum\limits_j g_j^2\rho_j(s)}
\label{bw1}
\ee
where $M=M_{BW}$ is called Breit-Wigner mass. For $\sum_j
g_j^2\rho_j(s)$ replaced by $M\Gamma$, we obtain the
non-relativistic Breit-Wigner amplitude. The pole position is
defined as zero of the amplitude denominator in the complex plane
\be
M^2-s-i\sum\limits_j g_j^2\rho_j(s)=0
\ee
and the partial width $\Gamma_a$ at $s=M^2$ (at the BW mass) is
defined as
\be
M\Gamma_a=g^2_a\rho_a(M^2)
\label{bww}
\ee
The helicity-dependent amplitude for photoproduction of the final
state $b$ can be written as
\be
a^h_b(s)=\frac{A_{BW}^h g_b}{M^2-s-i\sum\limits_j g_j^2\rho_j(s)}\,,
\ee
where $A_{BW}^h$ are photo-production couplings e.g. helicity
couplings in the helicity basis.

In general, the amplitude contains not only one resonance. In case
of several resonances (index $\alpha$), the $K$ matrix can be
written as:
\be
K_{ab}=\sum\limits_\alpha \frac{g^{\alpha}_a
g^{\alpha}_b}{M_\alpha^2-s}+f_{ab}\,.
\ee
Here the background terms $f_{ab}$ are added: they may be arbitrary
functions of $s$ and describe non-resonant transitions from the
initial to the final state.

The position of the pole $(M_{\rm pole}-i\frac 12 \Gamma_{\rm
pole})$ can be found by calculation of the zeros of the denominator
of a K-matrix amplitude in the complex s-plane
\cite{Anisovich:2007bq}
\begin{equation}
\label{pole}det(I-i\rho K)\prod\limits_{\alpha}(M_\alpha^2-s)=0.
\end{equation}
We define the residues for the transition amplitude by the contour
integral of the amplitude around the pole position in the energy
$(\sqrt s)$ plane to
\begin{eqnarray}
Res(a\to b)&=&\int\limits_{o}\frac{d\sqrt s}{2\pi
i}\sqrt{\rho_a}A_{ab}(s)\sqrt{\rho_b}\nonumber\label{transamp}\\
&=&\frac{1}{2M_p}\sqrt{\rho_a(M_p^2)}g^r_a\,g^r_b\sqrt{\rho_b(M_p^2)}\,.
\end{eqnarray}
Here $M_p$ is the position of the pole (complex number) and $g^r_a$
are pole couplings.
%\ee
The elastic pole residue is defined as
 \be
Res(\pi N\to N\pi)=\frac{1}{2M_p}
(g^r_{N\pi})^2\,\rho_{N\pi}(M_p^2)\,
\label{residuum}
 \ee
In the pole position one has a full factorization of the amplitude:
\be
Res^2(a\to b)=Res(a\to a)\times Res(b\to b)
\ee

The helicity-dependent amplitude for photoproduction of the final
state $b$ is calculated in the framework of P-vector approach:
\be
a^h_{b}=P^h_a(I-i\rho K)_{jb}^{-1}
\ee
where:
\be
P^h_a=\sum\limits_\alpha \frac{A^h_{\alpha}
g^{\alpha}_a}{M_\alpha^2-s}+F_{a}\,.
\ee
and $A^h_{\alpha}$ is photo-coupling of the K-matrix pole $\alpha$
and $F_{a}$ is a non-resonant transition. In the resonance pole the
photo-couplings are defined as:
\be
A^h g^{r}_b=\int\limits_{o}\frac{ds}{2\pi i}\;a^h_b(s)\,.
\label{helicity}
\ee
The helicity amplitudes $A^{1/2}$, $A^{3/2}$ (photo-couplings in the
helicity basis), the coupling elastic residues, and the residues of
the transition amplitudes are complex numbers. They become real and
coincide with the conventional helicity amplitudes $A^{1/2}$,
$A^{3/2}$, to half the elastic width $\Gamma_{N\pi}/2$, and to the
channel coupling $\frac12\sqrt{\Gamma_{\rm i}\Gamma_{\rm f}}$ if a
Breit-Wigner amplitude with constant width is used.

The elastic residue, which is proportional to
$(g^r_{N\pi})^2\rho_{N\pi}(M_p^2)$, defines $g^r_{N\pi}$ up to a
sign. This may lead to ambiguities if the phase is not properly
defined: assume the phase of elastic residue would be
$(180\pm\epsilon)^\circ$ in two analyses. Due to eq.
(\ref{helicity}), the phase of the helicity amplitude depends on
this definition. Since the phases of the elastic pole residue of
most resonances are negative, we define in the case of elastic
residues with a negative real part the phase of $g^r_{N\pi}$
clockwise.

In this article we also give some quantities which are related to
properties of a relativistic Breit-Wigner amplitude. We define the
Breit-Wigner amplitude by
\be
A_{ab}=\frac{f^2g^r_a g^r_b}{M_{BW}^2-s-if^2\sum\limits_a
|g^r_a|^2\rho_a(s)}
\label{bw}
\ee
where $M_{BW}$ and scaling factor $f$ are calculated to reproduce
exactly the pole position of the resonance. For a true Breit-Wigner
amplitude, $f=1$, and the definition in eq. (\ref{bw}) coincides
with the one in eq. (\ref{bw1}). In the case of a very fast growing
phase volume, the Breit-Wigner mass and width can shift from the
pole position by a large amount. For example, the Breit-Wigner mass
of the Roper resonance is 60-80\,MeV higher than the pole position
and its Breit-Wigner width exceeds the pole width by about 150\,MeV.
In the 1600-1700\,MeV region, the large phase volume leads to a very
large Breit-Wigner widths and an appreciable shift in mass from the
pole position (see for example \cite{Thoma:2007bm}) if the $\rho N$,
$\Delta \pi$ (with large $L$), and $D_{15}(1520)\pi$ decay modes are
taken into account explicitly. The visible width, e.g. in the $N\pi$
invariant mass spectrum, remains similar to the Breit-Wigner width.
Clearly, the large phase volume effects are highly model dependent
and possibly, they are artifacts of the formalism. We therefore
decided to extract the Breit-Wigner parameters of resonances above
the Roper resonance by approximating the phase volumes for the three
body channels in eq.(\ref{bw}) as $\pi N$ phase volume for the
respective partial wave. This procedure conserves the branching
ratio between three particle and $\pi N$ channels at the resonance
position and at the Breit-Wigner mass.

The Breit-Wigner helicity amplitude is defined as:
\be
a^h_{a}=\frac{A^h_{BW}fg^r_b}{M_{BW}^2-s-if^2\sum\limits_a
|g^r_a|^2\rho_a(s)}\,,
\label{helibw}
\ee
where $A^h_{BW}$ is calculated to reproduce the pion
photo-production residues in the pole. In general this quantum is a
complex number. However, for majority of resonances its phase
deviates only little from 0 or 180 degrees.

\section{Properties of baryon resonances}
On the subsequent pages we present properties of nucleon and
$\Delta$ resonances determined in this work. We give pole
parameters: pole position (eq. \ref{pole}), the complex helicity
amplitudes  $A^{1/2}$ and $A^{3/2}$ (eq. \ref{helicity}), the
elastic pole residue (eq. \ref{residuum}) and residues for hadronic
transition amplitudes (eq. \ref{transamp}).

The Tables also give properties of a relativistic Breit-Wigner
amplitude (eq. \ref{bw}), its helicity amplitudes (eq.
\ref{helibw}), partial decay widths (eq. \ref{bww}), and branching
ratios for the decay into channel $a$ by $\Gamma_a/\Gamma$.

A large number of resonances is required to achieve a good
description of all data sets. These resonances couple to a variety
of different decay modes. The optimum set of parameters is
determined in fits to the data of
Tables~\ref{piN_data_table}-\ref{chisquare1}. The fit minimizes the
total log likelihood defined by
\be
 -\ln {\cal L}_{\rm tot}= ( \frac 12\sum w_i\chi^2_i-\sum w_i\ln{\cal L}_i ) \ \frac{\sum
N_i}{\sum w_i N_i} \label{likeli}
\ee
where the summation over binned data contributes to the $\chi^2$
while unbinned data contribute to the likelihoods ${\cal L}_i$. Data
with $p\pi^0\pi^0$ and $p\pi^0\eta$ in the final state - except
those taken with polarized photons - are fitted event by event in
order to take into account all possible correlations between the
variables. For convenience of the reader, we quote differences in
fit quality as $\chi^2$ difference, with $\Delta\chi^2=-2\Delta{\cal
L}_{\rm tot}$. For new data, the weight is increased from $w_i=1$
until a visually acceptable fit is reached. Without weights,
low-statistics data e.g. on polarization variables may be reproduced
unsatisfactorily without significant deterioration of the total
${\cal L}_{\rm tot}$. The likelihood function is normalized to avoid
an artificial increase in statistics by the weighting factors.

Due to the incomplete data base with few double polarization
observables only, the solution of the partial wave analysis is not
unequivocal. Depending on the number of poles in the different
partial waves and depending on start values of the fit, different
minima of similar $\chi^2$ are reached. However, most parameters are
stable, only a few parameters undergo substantial changes. The
solutions which have converged to minima of similar depth are
stored; from the distribution of the fit results, typically more
than ten, the mean value and the error is deduced. As error we
assume the half-width of the distribution. In some cases, solutions
exist with a distinct minimum forming a new class of results, and
leading to a new set of parameters. Often, they cluster into two
main solutions, called BG2011-01 and BG2011-02. The most significant
difference can be found in the $1/2(3/2^+)$ wave where BG2011-02
finds two close-by resonances: $N(1900)3/2^+$, present in both types
of solutions with slightly different parameters, and $N(1975)3/2^+$,
present only in BG2011-02. Here, we give the properties of
$N(1900)3/2^+$ only. Sizable differences between the BG2011-01 and
BG2011-02 solutions are also observed in the $3/2^-$ (in particular
for $N(1700)3/2^-$), $5/2^+$ and $7/2^+$ wave. The different
solutions are discussed explicitly in \cite{Anisovich:2011ye}. Here,
we give errors which cover both solutions. The two solutions give
similar properties for $N(1880)1/2^+$ except for its helicity
amplitude. Here, we list both solutions in the Tables.

The 1700\,MeV region is complicated due to the presence of two
important thresholds, $N(1520)3/2^-\pi$ and $\Sigma K$.\\\noindent
$N(1520)3/2^-\pi$ in S-wave gives $3/2^+$ quantum numbers; in
\begin{table}[pb]
\caption{\label{rpp} (Next pages) Summary of results of the BnGa
partial wave analysis. The first blocks give quantities related to
the pole of the resonance, the second blocks give Breit-Wigner
parameters.}
\end{table}
\clearpage
\begin{table*}
\begin{tabular}{cc}
\hspace{-2mm}\begin{tabular}{lrlr} \multicolumn{4}{l}{\boldmath
\fbox{\fbox{$N(1440)\frac12^+$}}\unboldmath\qquad
or \quad $N(1440)P_{11}$}  \\[-0.7ex]
&&&\\\hline\hline\\[-1.5ex]
\multicolumn{4}{l}{$N(1440)\frac12^+$ pole parameters (MeV) }\\[0.3ex]\hline\\[-2ex]
$M_{\rm pole}$ &\hspace{15mm}1370\er4& $\Gamma_{\rm pole}$&\hspace{16mm}190\er7\\
\multicolumn{2}{l}{Elastic pole residue \hfill 48\er3}& Phase &-(78\er4)$^\circ$ \\
\multicolumn{2}{l}{Residue $\pi N\to N\sigma$ \hfill  20\er5 }& Phase &-(135\er7)$^\circ$ \\
\multicolumn{2}{l}{Residue $\pi N\to \Delta\pi$ \hfill  26\er3} & Phase & (40\er5)$^\circ$ \\[0.3ex]\hline\\[-2ex]
\multicolumn{2}{l}{$A^{1/2}$ (\,GeV$^{-\frac12}$)\hfill 0.044\er0.007}& Phase &(142\er5)$^\circ$\\
 & &  &\\
\hline\hline\\[-1.5ex]
\multicolumn{4}{l}{$N(1440)\frac12^+$ Breit-Wigner parameters (MeV)
}\\[0.3ex]\hline\\[-2ex]
$M_{\rm BW}$& 1430\er8 & $\Gamma_{\rm BW}$&365\er35 \\
Br($\pi N$)   &  62\er3\% \\
Br$(N\sigma$) &  17\er7\% &
Br($\Delta\pi$) &  21\er8\% \\[0.3ex]\hline\\[-2ex]
\multicolumn{2}{l}{$A_{BW}^{1/2}$ (\,GeV$^{-\frac12}$)\hfill
-0.061\er0.008}&& \\
\hline\hline\\
\multicolumn{4}{l}{\boldmath
\fbox{\fbox{$N(1535)\frac12^-$}}\unboldmath\qquad
or \quad $N(1535)S_{11}$}  \\[-0.7ex]
&&&\\\hline\hline\\[-1.5ex]
\multicolumn{4}{l}{$N(1535)\frac12^-$ pole parameters (MeV) }\\[0.3ex]\hline\\[-2ex]
$M_{\rm pole}$ &\hfill 1501\er4& $\Gamma_{\rm pole}$&\hfill 134\er11\\
\multicolumn{2}{l}{Elastic pole residue \hfill 31\er4}& Phase &-(29\er5)$^\circ$ \\
\multicolumn{2}{l}{Residue $\pi N\to N\eta$ \hfill  29\er4 }& Phase &-(76\er5)$^\circ$ \\
\multicolumn{2}{l}{Residue $\pi N\to \Delta\pi$ \hfill  8\er3} & Phase & (145\er17)$^\circ$\\
& &  & \\[0.3ex]\hline\\[-2ex]
\multicolumn{2}{l}{$A^{1/2}$ (\,GeV$^{-\frac12}$)\hfill 0.116\er0.010}& Phase &(7\er6)$^\circ$\\
\hline\hline\\[-1.5ex]
\multicolumn{4}{l}{$N(1535)\frac12^-$ Breit-Wigner parameters (MeV)
}\\[0.3ex]\hline\\[-2ex]
$M_{\rm BW}$& 1519\er5 & $\Gamma_{\rm BW}$&128\er14 \\
Br($\pi N$)   &  54\er5\% \\
Br$(N\eta$) &  33\er5\% &
Br($\Delta\pi$) &  2.5\er1.5\% \\[0.3ex]\hline\\[-2ex]
\multicolumn{2}{l}{$A_{BW}^{1/2}$ (\,GeV$^{-\frac12}$)\hfill
0.105\er0.010}&& \\
\hline\hline\\
\multicolumn{4}{l}{\boldmath
\fbox{\fbox{$N(1675)\frac52^-$}}\unboldmath\qquad
or \quad $N(1675)D_{15}$}  \\[-0.7ex]
&&&\\\hline\hline\\[-1.5ex]
\multicolumn{4}{l}{$N(1675)\frac52^-$ pole parameters (MeV) }\\[0.3ex]\hline\\[-2ex]
$M_{\rm pole}$ &\hfill 1654\er4& $\Gamma_{\rm pole}$&\hfill 151\er5\\
\multicolumn{2}{l}{Elastic pole residue \hfill 28\er1}& Phase &-(26\er4)$^\circ$ \\
\multicolumn{2}{l}{Residue $\pi N\to \Delta\pi$ \hfill  25\er5} & Phase & (82\er10)$^\circ$\\
\multicolumn{2}{l}{Residue $\pi N\to N\sigma$ \hfill  11\er4 }& Phase &(132\er18)$^\circ$ \\
& &  & \\[0.3ex]\hline\\[-2ex]
\multicolumn{2}{l}{$A^{1/2}$ (\,GeV$^{-\frac12}$)\hfill 0.024\er0.003}& Phase &-(16\er5)$^\circ$\\
\multicolumn{2}{l}{$A^{3/2}$ (\,GeV$^{-\frac12}$)\hfill 0.026\er0.008}& Phase &-(19\er6)$^\circ$\\
\hline\hline\\[-1.5ex]
\multicolumn{4}{l}{$N(1675)\frac52^-$ Breit-Wigner parameters (MeV)
}\\[0.3ex]\hline\\[-2ex]
$M_{\rm BW}$& 1664\er5 & $\Gamma_{\rm BW}$&152\er7 \\
Br($N \pi$)   &  40\er3\% \\
Br($\Delta\pi$) &  33\er8\% &  Br$(N\sigma$) & 7\er3\%
\\[0.3ex]\hline\\[-2ex]
\multicolumn{2}{l}{$A_{BW}^{1/2}$ (\,GeV$^{-\frac12}$)\hfill
0.024\er0.003}& \multicolumn{2}{l}{$A_{BW}^{3/2}$
(\,GeV$^{-\frac12}$)\hfill
0.025\er0.007} \\
\hline\hline\\
\end{tabular}&
\hspace{-2mm}\begin{tabular}{lrlr} \multicolumn{4}{l}{\boldmath
\fbox{\fbox{$N(1520)\frac32^-$}}\unboldmath\qquad
or \quad $N(1520)D_{13}$}  \\[-0.7ex]
&&&\\\hline\hline\\[-1.5ex]
\multicolumn{4}{l}{$N(1520)\frac32^-$ pole parameters (MeV) }\\[0.3ex]\hline\\[-2ex]
$M_{\rm pole}$ &\hfill 1507\er3& $\Gamma_{\rm pole}$&\hfill 111\er5\\
\multicolumn{2}{l}{Elastic pole residue \hfill 36\er3}& Phase &-(14\er3)$^\circ$ \\
\multicolumn{2}{l}{Residue {$\pi N\to \Delta\pi_{L\!=\!0}$}\hfill 18\er4} & Phase & (150\er20)$^\circ$ \\
\multicolumn{2}{l}{Residue {$\pi N\to \Delta\pi_{L\!=\!2}$}\hfill 14\er3} & Phase & (100\er20)$^\circ$ \\[0.3ex]\hline\\[-2ex]
\multicolumn{2}{l}{$A^{1/2}$ (\,GeV$^{-\frac12}$)\hfill -0.021\er0.004}& Phase &(0\er5)$^\circ$\\
\multicolumn{2}{l}{$A^{3/2}$ (\,GeV$^{-\frac12}$)\hfill  0.132\er0.009}& Phase &(2\er4)$^\circ$\\
\hline\hline\\[-1.5ex]
\multicolumn{4}{l}{$N(1520)\frac32^-$ Breit-Wigner parameters (MeV) }\\[0.3ex]\hline\\[-2ex]
$M_{\rm BW}$& 1517\er 3 & $\Gamma_{\rm BW}$&114\er5 \\
Br($\pi N$)   &  62\er3\% \\
Br$(\Delta\pi_{L\!=\!0})$ &  19\er4\% &
Br$(\Delta\pi_{L\!=\!2})$&\hspace{-6mm}   9\er2\% \\[0.3ex]\hline\\[-2ex]
\multicolumn{2}{l}{$A_{BW}^{1/2}$
(\,GeV$^{-\frac12}$)\hfill-0.022\er0.004}&
\multicolumn{2}{l}{$A_{BW}^{3/2}$ (\,GeV$^{-\frac12}$)\hfill0.131\er0.010} \\
\hline\hline\\
\multicolumn{4}{l}{\boldmath
\fbox{\fbox{$N(1650)\frac12^-$}}\unboldmath\qquad
or \quad $N(1650)S_{11}$}  \\[-0.7ex]
&&&\\\hline\hline\\[-1.5ex]
\multicolumn{4}{l}{$N(1650)\frac12^-$ pole parameters (MeV) }\\[0.3ex]\hline\\[-2ex]
$M_{\rm pole}$ &\hspace{11mm}1647\er6& $\Gamma_{\rm pole}$&\hspace{11mm} 103\er8\\
\multicolumn{2}{l}{Elastic pole residue \hfill 24\er3}& Phase &-(75\er12)$^\circ$ \\
\multicolumn{2}{l}{Residue {$\pi N\to N\eta$}\hfill 15\er2} & Phase & (134\er10)$^\circ$ \\
\multicolumn{2}{l}{Residue {$\pi N\to \Lambda K$}\hfill 11\er3} &
Phase & (85\er9)$^\circ$ \\
\multicolumn{2}{l}{Residue {$\pi N\to \Delta\pi$}\hfill 12\er3} & Phase & -(30\er20)$^\circ$ \\[0.3ex]\hline\\[-2ex]
\multicolumn{2}{l}{$A^{1/2}$ (\,GeV$^{-\frac12}$)\hfill 0.033\er0.007}& Phase &-(9\er15)$^\circ$\\
\hline\hline\\[-1.5ex]
\multicolumn{4}{l}{$N(1650)\frac12^-$ Breit-Wigner parameters (MeV) }\\[0.3ex]\hline\\[-2ex]
$M_{\rm BW}$& 1651\er 6 & $\Gamma_{\rm BW}$&104\er10 \\
Br($N\pi $)   &  51\er4\% &Br($N\eta $)   &  18\er4\% \\
Br$(\Lambda K)$ &  10\er5\% &
Br$(\Delta\pi)$&\hspace{-6mm}   19\er6\% \\[0.3ex]\hline\\[-2ex]
\multicolumn{2}{l}{$A_{BW}^{1/2}$
(\,GeV$^{-\frac12}$)\hfill 0.033\er0.007}\\
\hline\hline\\
\multicolumn{4}{l}{\boldmath
\fbox{\fbox{$N(1680)\frac52^+$}}\unboldmath\qquad
or \quad $N(1680)F_{15}$}  \\[-0.7ex]
&&&\\\hline\hline\\[-1.5ex]
\multicolumn{4}{l}{$N(1680)\frac52^+$ pole parameters (MeV) }\\[0.3ex]\hline\\[-2ex]
$M_{\rm pole}$ &\hfill 1676\er6& $\Gamma_{\rm pole}$&\hfill 113\er4\\
\multicolumn{2}{l}{Elastic pole residue \hfill 43\er4}& Phase &-(2\er10)$^\circ$ \\
\multicolumn{2}{l}{Residue {$\pi N\to \Delta\pi_{L\!=\!1}$}\hfill 8\er3} & Phase & -(70\er45)$^\circ$ \\
\multicolumn{2}{l}{Residue {$\pi N\to \Delta\pi_{L\!=\!3}$}\hfill 13\er3} & Phase & (85\er15)$^\circ$ \\
\multicolumn{2}{l}{Residue {$\pi N\to N\sigma$}\hfill 14\er3} & Phase & -(56\er15)$^\circ$ \\[0.3ex]\hline\\[-2ex]
\multicolumn{2}{l}{$A^{1/2}$ (\,GeV$^{-\frac12}$)\hfill -0.013\er0.004}& Phase &-(25\er22)$^\circ$\\
\multicolumn{2}{l}{$A^{3/2}$ (\,GeV$^{-\frac12}$)\hfill 0.134\er0.005}& Phase &-(2\er4)$^\circ$\\
\hline\hline\\[-1.5ex]
\multicolumn{4}{l}{$N(1680)\frac52^+$ Breit-Wigner parameters (MeV) }\\[0.3ex]\hline\\[-2ex]
$M_{\rm BW}$& 1689\er 6 & $\Gamma_{\rm BW}$&118\er6 \\
Br($N\pi $)   &  64\er5\% &Br($N\sigma $)   &  14\er7\% \\
Br$(\Delta\pi_{L\!=\!1})$&\hspace{-6mm}   10\er3\% &
Br$(\Delta\pi_{L\!=\!3})$&\hspace{-6mm}   5\er3\%
\\[0.3ex]\hline\\[-2ex]
\multicolumn{2}{l}{$A_{BW}^{1/2}$
(\,GeV$^{-\frac12}$)\hfill-0.013\er0.003}&
\multicolumn{2}{l}{$A_{BW}^{3/2}$ (\,GeV$^{-\frac12}$)\hfill0.135\er0.006} \\
\hline\hline\\
\end{tabular}
\end{tabular}
\end{table*}
\clearpage
\begin{table*}
\begin{tabular}{cc}
\hspace{-2mm}\begin{tabular}{lrlr} \multicolumn{4}{l}{\boldmath
\fbox{\fbox{$N(1700)\frac32^-$}}\unboldmath\qquad
or \quad $N(1700)D_{13}$}  \\[-0.7ex]
&&&\\\hline\hline\\[-1.5ex]
\multicolumn{4}{l}{$N(1700)\frac32^-$ pole parameters (MeV) }\\[0.3ex]\hline\\[-2ex]
$M_{\rm pole}$ &\hspace{9mm}1770\er40& $\Gamma_{\rm pole}$&\hspace{10mm}420\er180\\
\multicolumn{2}{l}{Elastic pole residue \hfill 50\er40}& Phase &-(100\er40)$^\circ$ \\
\multicolumn{2}{l}{Residue $\pi N\to \Delta\pi_{L\!=\!0}$ \hfill  75\er50} & Phase & -(60\er40)$^\circ$ \\
\multicolumn{2}{l}{Residue $\pi N\to \Delta\pi_{L\!=\!2}$ \hfill  18\er12} & Phase & (90\er35)$^\circ$ \\[0.3ex]\hline\\[-2ex]
\multicolumn{2}{l}{$A^{1/2}$ (\,GeV$^{-\frac12}$)\hfill~0.044\er0.020}& Phase &(85\er45)$^\circ$\\
\multicolumn{2}{l}{$A^{3/2}$ (\,GeV$^{-\frac12}$)\hfill~-0.037\er0.012}& Phase &(0\er30)$^\circ$\\
\hline\hline\\[-1.5ex]
\multicolumn{4}{l}{$N(1700)\frac32^-$ Breit-Wigner parameters (MeV)
}\\[0.3ex]\hline\\[-2ex]
$M_{\rm BW}$& 1790\er40 & $\Gamma_{\rm BW}$&390\er140 \\
Br($\pi N$)   &  12\er5\% \\
Br($\Delta\pi_{L\!=\!0}$) & 72\er16\% &
Br($\Delta\pi_{L\!=\!2}$) &  5\er4\% \\[0.3ex]\hline\\[-2ex]
\multicolumn{2}{l}{$A_{BW}^{1/2}$ (\,GeV$^{-\frac12}$)\hfill
~0.041\er0.017}& \multicolumn{2}{l}{$A_{BW}^{3/2}$
(\,GeV$^{-\frac12}$)\hfill
-0.034\er0.013} \\
\hline\hline\\
\multicolumn{4}{l}{\boldmath
\fbox{\fbox{$N(1720)\frac32^+$}}\unboldmath\qquad
or \quad $N(1720)P_{13}$}  \\[-0.7ex]
&&&\\\hline\hline\\[-1.5ex]
\multicolumn{4}{l}{$N(1720)\frac32^+$ pole parameters (MeV) }\\[0.3ex]\hline\\[-2ex]
$M_{\rm pole}$ &\hfill 1660\er30& $\Gamma_{\rm pole}$&\hfill 450\er100\\
\multicolumn{2}{l}{Elastic pole residue \hfill 22\er8}& Phase &-(115\er30)$^\circ$ \\
\multicolumn{2}{l}{Residue $\pi N\to N\eta$ \hfill  7\er5 }& Phase & not defined\\
\multicolumn{2}{l}{Residue $\pi N\to \Lambda K$ \hfill  14\er10 }& Phase &-(150\er45)$^\circ$ \\
\multicolumn{2}{l}{Residue $\pi N\to \Delta\pi_{L\!=\!1}$ \hfill  64\er25} & Phase & (80\er40)$^\circ$\\
\multicolumn{2}{l}{Residue $\pi N\to \Delta\pi_{L\!=\!3}$ \hfill
8\er8} & Phase & not defined
\\[0.3ex]\hline\\[-2ex]
\multicolumn{2}{l}{$A^{1/2}$ (\,GeV$^{-\frac12}$)\hfill 0.110\er0.045}& Phase &(0\er40)$^\circ$\\
\multicolumn{2}{l}{$A^{3/2}$ (\,GeV$^{-\frac12}$)\hfill 0.150\er0.035}& Phase &(65\er35)$^\circ$\\
\hline\hline\\[-1.5ex]
\multicolumn{4}{l}{$N(1720)\frac32^+$ Breit-Wigner parameters (MeV)
}\\[0.3ex]\hline\\[-2ex]
$M_{\rm BW}$& 1690$^{+70}_{-35}$ & $\Gamma_{\rm BW}$&420\er100 \\
Br($N\pi $)   &  10\er5\% &
Br$(N\eta$) &  3\er2\% \\
Br($\Delta\pi_{L\!=\!1}$) &  75\er15\% &
Br($\Delta\pi_{L\!=\!3}$) &  2\er2\% \\[0.3ex]\hline\\[-2ex]
\multicolumn{2}{l}{$A_{BW}^{1/2}$ (\,GeV$^{-\frac12}$)\hfill
0.110\er0.045}& \multicolumn{2}{l}{$A_{BW}^{3/2}$
(\,GeV$^{-\frac12}$)\hfill
0.150\er0.030} \\
\hline\hline\\
\multicolumn{4}{l}{\boldmath
\fbox{\fbox{$N(1875)\frac32^-$}}\unboldmath\qquad
or \quad $N(1875)D_{13}$}  \\[-0.7ex]
&&&\\\hline\hline\\[-1.5ex]
\multicolumn{4}{l}{$N(1875)\frac32^-$ pole parameters (MeV) }\\[0.3ex]\hline\\[-2ex]
$M_{\rm pole}$ &\hfill 1860\er25& $\Gamma_{\rm pole}$&\hfill 200\er20\\
\multicolumn{2}{l}{Elastic pole residue \hfill 2.5\er1.0}& Phase &not defined \\
\multicolumn{2}{l}{Residue $\pi N\to \Lambda K$ \hfill  1.5\er1.0} & Phase & not defined\\
\multicolumn{2}{l}{Residue $\pi N\to \Sigma K$ \hfill  5\er3} & Phase & not defined\\
\multicolumn{2}{l}{Residue $\pi N\to N\sigma$ \hfill  8\er3 }& Phase &-(170\er65)$^\circ$ \\
[0.3ex]\hline\\[-2ex]
\multicolumn{2}{l}{$A^{1/2}$ (\,GeV$^{-\frac12}$)\hfill 0.018\er0.008}& Phase &-(100\er60)$^\circ$\\
\multicolumn{2}{l}{$A^{3/2}$ (\,GeV$^{-\frac12}$)\hfill 0.010\er0.004}& Phase &~(180\er30)$^\circ$\\
\hline\hline\\[-1.5ex]
\multicolumn{4}{l}{$N(1875)\frac32^-$ Breit-Wigner parameters (MeV)
}\\[0.3ex]\hline\\[-2ex]
$M_{\rm BW}$& 1880\er20 & $\Gamma_{\rm BW}$&200\er25 \\
Br($N \pi$)   &  3\er2\% &Br($N \eta$)   &  5\er2\%\\
Br($\Lambda K$) &  4\er2\% &  Br$(\Sigma K$) & 15\er8\% \\
Br$(N\sigma$) & 60\er12\% &&
\\[0.3ex]\hline\\[-2ex]
\multicolumn{2}{l}{$A_{BW}^{1/2}$ (\,GeV$^{-\frac12}$)\hfill
0.018\er0.010}& \multicolumn{2}{l}{$A_{BW}^{3/2}$
(\,GeV$^{-\frac12}$)\hfill
-0.009\er0.005} \\
\hline\hline\\
\end{tabular} &
\hspace{-2mm}\begin{tabular}{lrlr} \multicolumn{4}{l}{\boldmath
\fbox{\fbox{$N(1710)\frac12^+$}}\unboldmath\qquad
or \quad $N(1710)P_{11}$}  \\[-0.7ex]
&&&\\\hline\hline\\[-1.5ex]
\multicolumn{4}{l}{$N(1710)\frac12^+$ pole parameters (MeV) }\\[0.3ex]\hline\\[-2ex]
$M_{\rm pole}$ &\hspace{12mm} 1687\er17& $\Gamma_{\rm pole}$&\hspace{12mm} 200\er25\\
\multicolumn{2}{l}{Elastic pole residue \hfill 6\er4}& Phase &(120\er70)$^\circ$ \\
\multicolumn{2}{l}{Residue {$\pi N\to N\eta$}\hfill 11\er4} & Phase & (0\er45)$^\circ$ \\
\multicolumn{2}{l}{Residue {$\pi N\to \Lambda K $}\hfill 17\er7} & Phase &-(110\er20)$^\circ$ \\[0.3ex]\hline\\[-2ex]
\multicolumn{2}{l}{$A^{1/2}$ (\,GeV$^{-\frac12}$)\hfill 0.055\er0.018}& Phase &-(10\er65)$^\circ$\\
~\\
\hline\hline\\[-1.5ex]
\multicolumn{4}{l}{$N(1710)\frac12^+$ Breit-Wigner parameters (MeV) }\\[0.3ex]\hline\\[-2ex]
$M_{\rm BW}$& 1710\er 20 & $\Gamma_{\rm BW}$&200\er18 \\
Br($N\pi$)   &  5\er4\% &Br($N\eta$)   &  17\er10\%\\
Br($\Lambda K$) &  23\er7\% &
 \\[0.3ex]\hline\\[-2ex]
\multicolumn{2}{l}{$A_{BW}^{1/2}$
(\,GeV$^{-\frac12}$)\hfill0.052\er0.015}&
\\
\hline\hline\\
\multicolumn{4}{l}{\boldmath
\fbox{\fbox{$N(1860)\frac52^+$}}\unboldmath\qquad
or \quad $N(1860)F_{15}$}  \\[-0.7ex]
&&&\\\hline\hline\\[-1.5ex]
\multicolumn{4}{l}{$N(1860)\frac52^+$ pole parameters (MeV) }\\[0.3ex]\hline\\[-2ex]
$M_{\rm pole}$ &\hfill $1830^{+120}_{-\ 60}$& $\Gamma_{\rm pole}$&\hfill $250^{+150}_{-\ 50}$\\
\multicolumn{2}{l}{Elastic pole residue \hfill 50\er20}& Phase &-(80\er40)$^\circ$ \\
&&&\\&&&\\[0.3ex]\hline\\[-2ex]
\multicolumn{2}{l}{$A^{1/2}$ (\,GeV$^{-\frac12}$)\hfill 0.020\er0.012}& Phase &(120\er50)$^\circ$\\
\multicolumn{2}{l}{$A^{3/2}$ (\,GeV$^{-\frac12}$)\hfill 0.050\er0.020}& Phase &-(80\er60)$^\circ$\\
\hline\hline\\[-1.5ex]
\multicolumn{4}{l}{$N(1860)\frac52^+$ Breit-Wigner parameters (MeV) }\\[0.3ex]\hline\\[-2ex]
$M_{\rm BW}$& $1860^{+120}_{-\ 60}$ & $\Gamma_{\rm BW}$& $270^{+140}_{-\ 50}$\\
Br($N\pi $)   &  20\er6\% & \\
&&& \\[0.3ex]\hline\\[-2ex]
\multicolumn{2}{l}{$A_{BW}^{1/2}$
(\,GeV$^{-\frac12}$)\hfill-0.019\er0.011}&
\multicolumn{2}{l}{$A_{BW}^{3/2}$ (\,GeV$^{-\frac12}$)\hfill0.048\er0.018} \\
\hline\hline\\
\multicolumn{4}{l}{\boldmath\fbox{\fbox{$N(1880)\frac12^+$}}\unboldmath\qquad
or \quad $N(1880)P_{11}$}\\[-0.7ex]
&&&\\\hline\hline\\[-1.5ex]
\multicolumn{4}{l}{$N(1880)\frac12^+$ pole parameters (MeV) }\\[0.3ex]\hline\\[-2ex]
$M_{\rm pole}$ &\hfill 1860\er35& $\Gamma_{\rm pole}$&\hfill 250\er 70\\
\multicolumn{2}{l}{Elastic pole residue \hfill 6\er4}& Phase &(80\er65)$^\circ$ \\
\multicolumn{2}{l}{Residue $\pi N\!\to\! \eta N$  \hfill 13\er 8}& Phase &-(75\er55)$^\circ$ \\
\multicolumn{2}{l}{Residue $\pi N\!\to\! \Lambda K$\hfill 4\er 3}& Phase & (40\er40)$^\circ$ \\
\multicolumn{2}{l}{Residue $\pi N\!\to\! \Sigma K$ \hfill 13\er 7}& Phase &(95\er40)$^\circ$ \\
\multicolumn{2}{l}{Residue $\pi N\!\to\! \Delta \pi$ \hfill 27\er 13}& Phase &-(150\er50)$^\circ$ \\
[0.3ex]\hline\\[-2ex]
\multicolumn{3}{l}{$A^{1/2}$ (\,GeV$^{-\frac12}$)\hfill 0.014\er0.003$^{(01)}$}  Phase & -(130\er60)$^\circ$\\
\multicolumn{3}{l}{$A^{1/2}$ (\,GeV$^{-\frac12}$)\hfill 0.036\er0.012$^{(02)}$}  Phase &  (15\er20)$^\circ$\\
\hline\hline\\[-1.5ex]
\multicolumn{4}{l}{$N(1880)\frac12^+$ Breit-Wigner parameters (MeV) }\\
[0.3ex]\hline\\[-2ex]
$M_{\rm BW}$& 1870\er35 & $\Gamma_{\rm BW}$& 235\er 65\\
Br($\pi N$)    &  5\er 3\% &
Br($\eta N$)   &  $25^{+30}_{-20}$\% \\
Br($\Lambda K$) &  2\er 1\% &
Br($\Sigma K$)  &  17\er7\% \\
Br($\Delta\pi$)  &  29\er12\% \\
[0.3ex]\hline\\[-2ex]
\multicolumn{4}{l}{$A_{BW}^{1/2}$ (\,GeV$^{-\frac12}$)\hfill -0.013\er0.003}$^{(01)}$  \\
\multicolumn{4}{l}{$A_{BW}^{1/2}$ (\,GeV$^{-\frac12}$)\hfill ~0.034\er0.011}$^{(02)}$  \\
\hline\hline\\
\end{tabular}
\end{tabular}
\end{table*}
\clearpage
\begin{table*}
\begin{tabular}{cc}
\hspace{-2mm}\begin{tabular}{lrlr}\multicolumn{4}{l}{\boldmath
\fbox{\fbox{$N(1895)\frac12^-$}}\unboldmath\qquad
or \quad $N(1895)S_{11}$}\\[-0.7ex]
&&&\\\hline\hline\\[-1.5ex]
\multicolumn{4}{l}{$N(1895)\frac12^-$ pole parameters (MeV) }\\[0.3ex]\hline\\[-2ex]
$M_{\rm pole}$ &\hspace{14mm}1900\er15& $\Gamma_{\rm pole}$&\hspace{14mm}$90^{+30}_{-15}$\\
\multicolumn{2}{l}{Elastic pole residue \hfill 1\er1}& Phase &not defined \\
\multicolumn{2}{l}{Residue $\pi N\!\to\! \eta N$  \hfill 3\er 2}& Phase &(40\er20)$^\circ$ \\
\multicolumn{2}{l}{Residue $\pi N\!\to\! K\Lambda$\hfill  2\er1}& Phase &-(90\er30)$^\circ$ \\
\multicolumn{2}{l}{Residue $\pi N\!\to\! K\Sigma$ \hfill  3\er2}& Phase &(40\er30)$^\circ$ \\
[0.3ex]\hline\\[-2ex]
\multicolumn{2}{l}{$A^{1/2}$ (\,GeV$^{-\frac12}$)\hfill 0.012\er0.006} & Phase & (120\er50)$^\circ$\\
&&&\\\hline\hline\\[-1.5ex]
\multicolumn{4}{l}{$N(1895)\frac12^-$ Breit-Wigner parameters (MeV) }\\
[0.3ex]\hline\\[-2ex]
$M_{\rm BW}$& 1895\er15 & $\Gamma_{\rm BW}$& $90^{+30}_{-15}$\\
Br($\pi N$)    &  2\er1\% &
Br($\eta N$)   &  21\er6\% \\
Br($K\Lambda$) &  18\er5\% &
Br($K\Sigma$)  &  13\er7\% \\[0.3ex]\hline\\[-2ex]
\multicolumn{2}{l}{$A_{BW}^{1/2}$ (\,GeV$^{-\frac12}$)\hfill-0.011\er0.006} && \\
\hline\hline\\
\multicolumn{4}{l}{\boldmath
\fbox{\fbox{$N(1990)\frac72^+$}}\unboldmath\qquad
or \quad $N(1990)F_{17}$}\\[-0.7ex]
&&&\\\hline\hline\\[-1.5ex]
\multicolumn{4}{l}{$N(1990)\frac72^+$ pole parameters (MeV) }\\[0.3ex]\hline\\[-2ex]
$M_{\rm pole}$ &\hfill 2030\er65& $\Gamma_{\rm pole}$&\hfill 240\er60\\
\multicolumn{2}{l}{Elastic pole residue \hfill 2\er1}& Phase &  (125\er65)$^\circ$ \\
&&&\\[0.3ex]\hline\\[-2ex]
\multicolumn{2}{l}{$A^{1/2}$ (\,GeV$^{-\frac12}$)\hfill  0.042\er0.014} & Phase &-(30\er20)$^\circ$\\
\multicolumn{2}{l}{$A^{3/2}$ (\,GeV$^{-\frac12}$)\hfill  0.058\er0.012} & Phase &-(35\er25)$^\circ$\\
\hline\hline\\[-1.5ex]
\multicolumn{4}{l}{$N(1990)\frac72^+$ Breit-Wigner parameters (MeV) }\\[0.3ex]\hline\\[-2ex]
$M_{\rm BW}$& 2060\er65 & $\Gamma_{\rm BW}$&240\er50 \\
Br($\pi N$)   &2\er1\% & &\\
[0.3ex]\hline\\[-2ex]
\multicolumn{2}{l}{$A_{BW}^{1/2}$ (\,GeV$^{-\frac12}$)\hfill
0.040\er 0.012} &
\multicolumn{2}{l}{$A_{BW}^{3/2}$ (\,GeV$^{-\frac12}$)\hfill  0.057\er 0.012} \\
\hline\hline\\
\multicolumn{4}{l}{\boldmath
\fbox{\fbox{$N(2060)\frac52^-$}}\unboldmath\qquad
or \quad $N(2060)D_{15}$}\\[-0.7ex]
&&&\\\hline\hline\\[-1.5ex]
\multicolumn{4}{l}{$N(2060)\frac52^-$ pole parameters (MeV) }\\[0.3ex]\hline\\[-2ex]
$M_{\rm pole}$ &\hfill 2040\er15& $\Gamma_{\rm pole}$&\hfill 390\er25\\
\multicolumn{2}{l}{Elastic pole residue \hfill 19\er5}& Phase & -(125\er20)$^\circ$ \\
\multicolumn{2}{l}{Residue {$\pi N\!\to\!\eta N$} \hfill 12\er6} & Phase &(40\er25)$^\circ$ \\
\multicolumn{2}{l}{Residue {$\pi N\!\to\!K\Lambda$}\hfill 1\er0.5} & Phase & not defined \\
\multicolumn{2}{l}{Residue {$\pi N\!\to\!K\Sigma$} \hfill 7\er4} & Phase & -(70\er30)$^\circ$ \\
[0.3ex]\hline\\[-2ex]
\multicolumn{2}{l}{$A^{1/2}$ (\,GeV$^{-\frac12}$)\hfill  0.065\er0.012} & Phase & (15\er8)$^\circ$\\
\multicolumn{2}{l}{$A^{3/2}$ (\,GeV$^{-\frac12}$)\hfill  $0.055^{+15}_{-35}$} & Phase & (15\er10)$^\circ$\\
\hline\hline\\[-1.5ex]
\multicolumn{4}{l}{$N(2060)\frac52^-$ Breit-Wigner parameters (MeV) }\\[0.3ex]\hline\\[-2ex]
$M_{\rm BW}$& 2060\er15 & $\Gamma_{\rm BW}$&375\er25 \\
Br($\pi N$)   &8\er2\% &
Br($\eta N$) &  4\er2\%  \\
Br($K\Sigma$) &  3\er2\%  &
  &   \\
[0.3ex]\hline\\[-2ex]
\multicolumn{2}{l}{$A_{BW}^{1/2}$ (\,GeV$^{-\frac12}$)\hfill
0.067\er 0.015} &
\multicolumn{2}{l}{$A_{BW}^{3/2}$ (\,GeV$^{-\frac12}$)\hfill  0.055\er 0.020} \\
\hline\hline\\
\end{tabular}&
\begin{tabular}{lrlr}
\multicolumn{4}{l}{\boldmath
\fbox{\fbox{$N(1900)\frac32^+$}}\unboldmath\qquad
or \quad $N(1900)P_{13}$}\\[-0.7ex]
&&&\\\hline\hline\\[-1.5ex]
\multicolumn{4}{l}{$N(1900)\frac32^+$ pole parameters (MeV) }\\[0.3ex]\hline\\[-2ex]
$M_{\rm pole}$ &\hfill 1900\er30& $\Gamma_{\rm pole}$&\hfill $260^{+100}_{-60}$\\
\multicolumn{2}{l}{Elastic pole residue \hfill 3\er2}& Phase & (10\er35)$^\circ$ \\
\multicolumn{2}{l}{Residue {$\pi N\!\to\!\eta N$} \hfill 6\er3} & Phase &(70\er60)$^\circ$ \\
\multicolumn{2}{l}{Residue {$\pi N\!\to\!K\Lambda$}\hfill 9\er5} & Phase & (135\er25)$^\circ$ \\
\multicolumn{2}{l}{Residue {$\pi N\!\to\!K\Sigma$} \hfill 5\er3} & Phase & (110\er30)$^\circ$ \\
[0.3ex]\hline\\[-2ex]
\multicolumn{2}{l}{$A^{1/2}$ (\,GeV$^{-\frac12}$)\hfill  0.026\er0.015} & Phase & (60\er40)$^\circ$\\
\multicolumn{2}{l}{$A^{3/2}$ (\,GeV$^{-\frac12}$)\hfill  0.060\er0.030} & Phase & (185\er60)$^\circ$\\
\hline\hline\\[-1.5ex]
\multicolumn{4}{l}{$N(1900)\frac32^+$ Breit-Wigner parameters (MeV) }\\[0.3ex]\hline\\[-2ex]
$M_{\rm BW}$& 1905\er30 & $\Gamma_{\rm BW}$&$250^{+120}_{-50}$ \\
Br($\pi N$)   &  3\er2\% &
Br($\eta N$) &  10\er4\%  \\
Br($K\Lambda$) &  16\er5\%  &
Br($K\Sigma$) &  5\er2\%  \\[0.3ex]\hline\\[-2ex]
\multicolumn{2}{l}{$A_{BW}^{1/2}$ (\,GeV$^{-\frac12}$)\hfill
0.026\er 0.015} &
\multicolumn{2}{l}{$A_{BW}^{3/2}$ (\,GeV$^{-\frac12}$)\hfill -0.065\er 0.030} \\
\hline\hline\\
\multicolumn{4}{l}{\boldmath
\fbox{\fbox{$N(2000)\frac52^+$}}\unboldmath\qquad
or \quad $N(2000)F_{15}$}\\[-0.7ex]
&&&\\\hline\hline\\[-1.5ex]
\multicolumn{4}{l}{$N(2000)\frac52^+$ pole parameters (MeV) }\\[0.3ex]\hline\\[-2ex]
$M_{\rm pole}$ &\hfill 2030\er110& $\Gamma_{\rm pole}$&\hfill 480\er100\\
\multicolumn{2}{l}{Elastic pole residue \hfill $35^{+80}_{-15}$}& Phase &-(100\er40)$^\circ$ \\
&&& \\
[0.3ex]\hline\\[-2ex]
\multicolumn{2}{l}{$A^{1/2}$ (\,GeV$^{-\frac12}$)\hfill  0.035\er0.015} & Phase & (15\er40)$^\circ$\\
\multicolumn{2}{l}{$A^{3/2}$ (\,GeV$^{-\frac12}$)\hfill  0.050\er0.014} & Phase &-(130\er40)$^\circ$\\
\hline\hline\\[-1.5ex]
\multicolumn{4}{l}{$N(2000)\frac52^+$ Breit-Wigner parameters (MeV) }\\[0.3ex]\hline\\[-2ex]
$M_{\rm BW}$& 2090\er120 & $\Gamma_{\rm BW}$&460\er100 \\
Br($\pi N$)   &  9\er4\% \\
[0.3ex]\hline\\[-2ex]
\multicolumn{2}{l}{$A_{BW}^{1/2}$ (\,GeV$^{-\frac12}$)\hfill
0.032\er 0.014} &
\multicolumn{2}{l}{$A_{BW}^{3/2}$ (\,GeV$^{-\frac12}$)\hfill  0.048\er 0.014} \\
\hline\hline\\
\multicolumn{4}{l}{\boldmath
\fbox{\fbox{$N(2150)\frac32^-$}}\unboldmath\qquad
or \quad $N(2150)D_{13}$}\\[-0.7ex]
&&&\\\hline\hline\\[-1.5ex]
\multicolumn{4}{l}{$N(2150)\frac32^-$ pole parameters (MeV) }\\[0.3ex]\hline\\[-2ex]
$M_{\rm pole}$ &\hfill 2110\er50& $\Gamma_{\rm pole}$&\hfill 340\er45\\
\multicolumn{2}{l}{Elastic pole residue \hfill 13\er3}& Phase & -(20\er10)$^\circ$ \\
\multicolumn{2}{l}{Residue {$\pi N\!\to\!K\Lambda$}\hfill 5\er2} & Phase & (100\er30)$^\circ$ \\
\multicolumn{2}{l}{Residue {$\pi N\!\to\!K\Sigma$} \hfill 3\er2} & Phase & -(50\er40)$^\circ$ \\
&&&\\[0.3ex]\hline\\[-2ex]
\multicolumn{2}{l}{$A^{1/2}$ (\,GeV$^{-\frac12}$)\hfill  0.125\er0.045} & Phase &-(55\er20)$^\circ$\\
\multicolumn{2}{l}{$A^{3/2}$ (\,GeV$^{-\frac12}$)\hfill  0.150\er0.060} & Phase &-(35\er15)$^\circ$\\
\hline\hline\\[-1.5ex]
\multicolumn{4}{l}{$N(2150)\frac32^-$ Breit-Wigner parameters (MeV) }\\[0.3ex]\hline\\[-2ex]
$M_{\rm BW}$& 2150\er60 & $\Gamma_{\rm BW}$&330\er45 \\
Br($\pi N$)     &6\er2\% &&\\
&&&\\[0.3ex]\hline\\[-2ex]
\multicolumn{2}{l}{$A_{BW}^{1/2}$ (\,GeV$^{-\frac12}$)\hfill
0.130\er 0.045} &
\multicolumn{2}{l}{$A_{BW}^{3/2}$ (\,GeV$^{-\frac12}$)\hfill  0.150\er 0.055} \\
\hline\hline\\
\end{tabular}
\end{tabular}
\end{table*}
\clearpage
\begin{table*}
\begin{tabular}{cc}
\hspace{-2mm}\begin{tabular}{lrlr}\multicolumn{4}{l}{\boldmath
\fbox{\fbox{$N(2190)\frac72^-$}}\unboldmath\qquad
or \quad $N(2190)G_{17}$}\\[-0.7ex]
&&&\\\hline\hline\\[-1.5ex]
\multicolumn{4}{l}{$N(2190)\frac72^-$ pole parameters (MeV) }\\[0.3ex]\hline\\[-2ex]
$M_{\rm pole}$ &\hspace{14mm}2150\er25& $\Gamma_{\rm pole}$&\hspace{11mm}330\er30\\
\multicolumn{2}{l}{Elastic pole residue \hfill 30\er5}& Phase & (30\er10)$^\circ$ \\
\multicolumn{2}{l}{Residue {$\pi N\!\to\!K\Lambda$}\hfill 4.9\er1.5} & Phase & (20\er15)$^\circ$ \\
[0.3ex]\hline\\[-2ex]
\multicolumn{2}{l}{$A^{1/2}$ (\,GeV$^{-\frac12}$)\hfill  0.063\er0.007} & Phase &-(170\er15)$^\circ$\\
\multicolumn{2}{l}{$A^{3/2}$ (\,GeV$^{-\frac12}$)\hfill  0.035\er0.020} & Phase & (25\er10)$^\circ$\\
\hline\hline\\[-1.5ex]
\multicolumn{4}{l}{$N(2190)\frac72^-$ Breit-Wigner parameters (MeV) }\\[0.3ex]\hline\\[-2ex]
$M_{\rm BW}$& 2180\er20 & $\Gamma_{\rm BW}$&335\er40 \\
Br($\pi N$)   & 16\er2\% & Br($K\Lambda$)& 0.5\er0.3\%  \\
[0.3ex]\hline\\[-2ex]
\multicolumn{2}{l}{$A_{BW}^{1/2}$ (\,GeV$^{-\frac12}$)\hfill
-0.065\er 0.008} &
\multicolumn{2}{l}{$A_{BW}^{3/2}$ (\,GeV$^{-\frac12}$)\hfill 0.035\er 0.017} \\
\hline\hline\\
\multicolumn{4}{l}{\boldmath
\fbox{\fbox{$N(2250)\frac92^-$}}\unboldmath\qquad
or \quad $N(2250)G_{19}$}\\[-0.7ex]
&&&\\\hline\hline\\[-1.5ex]
\multicolumn{4}{l}{$N(2250)\frac92^-$ pole parameters (MeV) }\\[0.3ex]\hline\\[-2ex]
$M_{\rm pole}$ &\hfill 2195\er45& $\Gamma_{\rm pole}$&\hfill 470\er50\\
\multicolumn{2}{l}{Elastic pole residue \hfill 26\er5}& Phase &-(38\er25)$^\circ$ \\
[0.3ex]\hline\\[-2ex]
\multicolumn{2}{l}{$A^{1/2}$ (\,GeV$^{-\frac12}$)\hfill  $<0.010$} & Phase &not defined\\
\multicolumn{2}{l}{$A^{3/2}$ (\,GeV$^{-\frac12}$)\hfill  $<0.010$} & Phase &not defined\\
\hline\hline\\[-1.5ex]
\multicolumn{4}{l}{$N(2250)\frac92^-$ Breit-Wigner parameters (MeV) }\\[0.3ex]\hline\\[-2ex]
$M_{\rm BW}$& 2280\er40 & $\Gamma_{\rm BW}$&520\er50 \\
Br($\pi N$)   & 12\er4\% \\
[0.3ex]\hline\\[-2ex]
\multicolumn{2}{l}{$|A_{BW}^{1/2}|$ (\,GeV$^{-\frac12}$)$<0.010$} &
\multicolumn{2}{l}{$|A_{BW}^{3/2}|$ (\,GeV$^{-\frac12}$)$<0.010$} \\
\hline\hline\\
\multicolumn{4}{l}{\boldmath
\fbox{\fbox{$\Delta(1232)\frac32^+$}}\unboldmath\qquad
or \quad $\Delta(1232)P_{33}$}  \\[-0.7ex]
&&&\\\hline\hline\\[-1.5ex]
\multicolumn{4}{l}{$\Delta(1232)\frac32^+$ pole parameters (MeV) }\\[0.3ex]\hline\\[-2ex]
$M_{\rm pole}$ &\hfill 1210.5\er1.0& $\Gamma_{\rm pole}$&\hfill 99\er2\\
\multicolumn{2}{l}{Elastic pole residue \hfill 51.6\er0.6}& Phase &-(46\er1)$^\circ$ \\
~ \\
~ \\
[0.3ex]\hline\\[-2ex]
\multicolumn{2}{l}{$A^{1/2}$ (\,GeV$^{-\frac12}$)\hfill-0.131\er0.0035}& Phase &-(19\er2)$^\circ$\\
\multicolumn{2}{l}{$A^{3/2}$ (\,GeV$^{-\frac12}$)\hfill-0.254\er0.0045}& Phase &-(9\er1)$^\circ$\\
\hline\hline\\[-1.5ex]
\multicolumn{4}{l}{$\Delta(1232)\frac32^-$ Breit-Wigner parameters
(MeV)
}\\[0.3ex]\hline\\[-2ex]
$M_{\rm BW}$& 1228\er2 & $\Gamma_{\rm BW}$&110\er3 \\
~\\
~\\
[0.3ex]\hline\\[-2ex]
\multicolumn{2}{l}{$A_{BW}^{1/2}$ (\,GeV$^{-\frac12}$)\hfill
-0.131\er0.004}&
\multicolumn{2}{l}{$A_{BW}^{3/2}$ (\,GeV$^{-\frac12}$)\hfill-0.254\er0.005} \\
\hline\hline\\
\end{tabular}&
\begin{tabular}{lrlr}
\multicolumn{4}{l}{\boldmath
\fbox{\fbox{$N(2220)\frac92^+$}}\unboldmath\qquad
or \quad $N(2220)H_{19}$}\\[-0.7ex]
&&&\\\hline\hline\\[-1.5ex]
\multicolumn{4}{l}{$N(2220)\frac92^+$ pole parameters (MeV) }\\[0.3ex]\hline\\[-2ex]
$M_{\rm pole}$ &\hfill 2150\er35& $\Gamma_{\rm pole}$&\hfill 440\er40\\
\multicolumn{2}{l}{Elastic pole residue \hfill 60\er12}& Phase &-(58\er12)$^\circ$ \\
~\\[0.3ex]\hline\\[-2ex]
\multicolumn{2}{l}{$A^{1/2}$ (\,GeV$^{-\frac12}$)\hfill  $<0.010$} & Phase &not defined\\
\multicolumn{2}{l}{$A^{3/2}$ (\,GeV$^{-\frac12}$)\hfill  $<0.010$} & Phase &not defined\\
\hline\hline\\[-1.5ex]
\multicolumn{4}{l}{$N(2220)\frac92^+$ Breit-Wigner parameters (MeV) }\\[0.3ex]\hline\\[-2ex]
$M_{\rm BW}$& 2200\er50 & $\Gamma_{\rm BW}$&480\er60 \\
Br($\pi N$)   & 24\er5\% \\[0.3ex]\hline\\[-2ex]
\multicolumn{2}{l}{$|A_{BW}^{1/2}|$ (\,GeV$^{-\frac12}$)$<0.010$} &
\multicolumn{2}{l}{$|A_{BW}^{3/2}|$ (\,GeV$^{-\frac12}$)$<0.010$} \\
\hline\hline\\[45.4ex]
\multicolumn{4}{l}{\boldmath
\fbox{\fbox{$\Delta(1600)\frac32^+$}}\unboldmath\qquad
or \quad $\Delta(1600)P_{33}$}  \\[-0.7ex]
&&&\\\hline\hline\\[-1.5ex]
\multicolumn{4}{l}{$\Delta(1600)\frac32^+$ pole parameters (MeV) }\\[0.3ex]\hline\\[-2ex]
$M_{\rm pole}$ &\hspace{10mm}1498\er25& $\Gamma_{\rm pole}$&\hspace{11mm}230\er50\\
\multicolumn{2}{l}{Elastic pole residue \hfill 11\er6}& Phase &-(160\er33)$^\circ$ \\
\multicolumn{2}{l}{Residue $\pi N\to \Delta\pi_{L\!=\!1}$\hfill 18\er15} & Phase & (154\er40)$^\circ$ \\
\multicolumn{2}{l}{Residue $\pi N\to \Delta\pi_{L\!=\!3}$\hfill 1\er1} & Phase & ~ \\
[0.3ex]\hline\\[-2ex]
\multicolumn{2}{l}{$A^{1/2}$ (\,GeV$^{-\frac12}$)\hfill~0.053\er0.010}& Phase &(130\er25)$^\circ$\\
\multicolumn{2}{l}{$A^{3/2}$ (\,GeV$^{-\frac12}$)\hfill~0.041\er0.011}& Phase &(165\er17)$^\circ$\\
\hline\hline\\[-1.5ex]
\multicolumn{4}{l}{$\Delta(1600)\frac32^+$ Breit-Wigner parameters (MeV) }\\[0.3ex]\hline\\[-2ex]
$M_{\rm BW}$& 1510\er 20 & $\Gamma_{\rm BW}$&220\er45 \\
Br($N\pi$)   &  12\er5\% \\
Br($\Delta\pi_{L\!=\!1})$&  78\er6\% &
Br($\Delta\pi_{L\!=\!3})$&  2\er2\% \\
[0.3ex]\hline\\[-2ex]
\multicolumn{2}{l}{$A_{BW}^{1/2}$
(\,GeV$^{-\frac12}$)\hfill-0.050\er0.009}&
\multicolumn{2}{l}{$A_{BW}^{3/2}$ (\,GeV$^{-\frac12}$)\hfill-0.040\er0.012} \\
\hline\hline\\
\end{tabular}
\end{tabular}
\end{table*}
\clearpage
\begin{table*}
\begin{tabular}{cc}
\hspace{-2mm}\begin{tabular}{lrlr} \multicolumn{4}{l}{\boldmath
\fbox{\fbox{$\Delta(1620)\frac12^-$}}\unboldmath\qquad
or \quad $\Delta(1620)S_{31}$}  \\[-0.7ex]
&&&\\\hline\hline\\[-1.5ex]
\multicolumn{4}{l}{$\Delta(1620)\frac12^-$ pole parameters (MeV) }\\[0.3ex]\hline\\[-2ex]
$M_{\rm pole}$ &\hspace{15mm}1597\er4& $\Gamma_{\rm pole}$&\hspace{12mm}130\er9\\
\multicolumn{2}{l}{Elastic pole residue \hfill 18\er2}& Phase &-(100\er5)$^\circ$ \\
\multicolumn{2}{l}{Residue $\pi N\to \Delta\pi$ \hfill  25\er5} &
Phase & -(85\er30)$^\circ$
\\[0.3ex]\hline\\[-2ex]
\multicolumn{2}{l}{$A^{1/2}$ (\,GeV$^{-\frac12}$)\hfill 0.052\er0.005}& Phase &-(9\er9)$^\circ$\\
[0.4ex]~\\
\hline\hline\\[-1.5ex]
\multicolumn{4}{l}{$\Delta(1620)\frac12^-$ Breit-Wigner parameters
(MeV)
}\\[0.3ex]\hline\\[-2ex]
$M_{\rm BW}$& 1600\er8 & $\Gamma_{\rm BW}$&130\er11 \\
Br($N\pi $)   &  28\er3\% &
Br($\Delta\pi$) &  60\er12\% \\
~\\[0.3ex]\hline\\[-2ex]
\multicolumn{2}{l}{$A_{BW}^{1/2}$ (\,GeV$^{-\frac12}$)\hfill
~0.052\er0.005}&& \\[0.4ex]
\hline\hline\\
\multicolumn{4}{l}{\boldmath
\fbox{\fbox{$\Delta(1900)\frac12^-$}}\unboldmath\qquad
or \quad $\Delta(1900)S_{31}$}  \\[-0.7ex]
&&&\\\hline\hline\\[-1.5ex]
\multicolumn{4}{l}{$\Delta(1900)\frac12^-$ pole parameters (MeV) }\\[0.3ex]\hline\\[-2ex]
$M_{\rm pole}$ &\hspace{12mm}1845\er25& $\Gamma_{\rm pole}$&\hspace{9mm}300\er45\\
\multicolumn{2}{l}{Elastic pole residue \hfill 10\er3}& Phase &-(125\er20)$^\circ$ \\
\multicolumn{2}{l}{Residue $\pi N\to \Sigma K$ \hfill  10\er3} & Phase &-(50\er30)$^\circ$\\
\multicolumn{2}{l}{Residue $\pi N\to \Delta\pi$ \hfill  15\er10} &
Phase &(110\er20)$^\circ$
\\[0.3ex]\hline\\[-2ex]
\multicolumn{2}{l}{$A^{1/2}$ (\,GeV$^{-\frac12}$)\hfill 0.059\er0.016}& Phase &(60\er25)$^\circ$\\
[0.3ex]~\\
\hline\hline\\[-1.5ex]
\multicolumn{4}{l}{$\Delta(1900)\frac12^-$ Breit-Wigner parameters
(MeV)
}\\[0.3ex]\hline\\[-2ex]
$M_{\rm BW}$& 1840\er30 & $\Gamma_{\rm BW}$&300\er45 \\
Br($N\pi $)   &  7\er3\% &Br($\Sigma K $)   &  5\er3\%\\
Br($\Delta\pi$) &  58\er25\% \\
[0.3ex]\hline\\[-2ex]
\multicolumn{2}{l}{$A_{BW}^{1/2}$ (\,GeV$^{-\frac12}$)\hfill
~0.059\er0.016}&& \\[0.3ex]
~\\
\hline\hline\\
\multicolumn{4}{l}{\boldmath
\fbox{\fbox{$\Delta(1910)\frac12^+$}}\unboldmath\qquad
or \quad $\Delta(1910)P_{31}$}  \\[-0.7ex]
&&&\\\hline\hline\\[-1.5ex]
\multicolumn{4}{l}{$\Delta(1910)\frac12^+$ pole parameters (MeV) }\\[0.3ex]\hline\\[-2ex]
$M_{\rm pole}$ &\hspace{12mm}1850\er40& $\Gamma_{\rm pole}$&\hspace{9mm}350\er45\\
\multicolumn{2}{l}{Elastic pole residue \hfill 24\er6}& Phase &-(145\er30)$^\circ$ \\
\multicolumn{2}{l}{Residue $\pi N\to \Sigma K$ \hfill  12\er4} & Phase & -(110\er30)$^\circ$\\
\multicolumn{2}{l}{Residue $\pi N\to \Delta\pi$ \hfill  30\er14} &
Phase & (95\er40)$^\circ$
\\[0.3ex]\hline\\[-2ex]
\multicolumn{2}{l}{$A^{1/2}$ (\,GeV$^{-\frac12}$)\hfill 0.023\er0.009}& Phase & (40\er90)$^\circ$\\
[0.4ex]~\\
\hline\hline\\[-1.5ex]
\multicolumn{4}{l}{$\Delta(1910)\frac12^-$ Breit-Wigner parameters
(MeV)
}\\[0.3ex]\hline\\[-2ex]
$M_{\rm BW}$& 1860\er40 & $\Gamma_{\rm BW}$&350\er55 \\
Br($N\pi $)   &  12\er3\% &Br($\Sigma K $)   &  9\er5\%\\
Br($\Delta\pi$) &  60\er20\%
~\\[0.3ex]\hline\\[-2ex]
\multicolumn{2}{l}{$A_{BW}^{1/2}$ (\,GeV$^{-\frac12}$)\hfill
~0.022\er0.009}&& \\[0.4ex]
~\\
\hline\hline\\
\end{tabular}&\hspace{-2mm}\begin{tabular}{lrlr}
\multicolumn{4}{l}{\boldmath
\fbox{\fbox{$\Delta(1700)\frac32^-$}}\unboldmath\qquad
or \quad $\Delta(1700)D_{33}$}  \\[-0.7ex]
&&&\\\hline\hline\\[-1.5ex]
\multicolumn{4}{l}{$\Delta(1700)\frac32^-$ pole parameters (MeV) }\\[0.3ex]\hline\\[-2ex]
$M_{\rm pole}$ &\hspace{11mm}1680\er10& $\Gamma_{\rm pole}$&\hspace{11mm}305\er15\\
\multicolumn{2}{l}{Elastic pole residue \hfill 42\er7}& Phase &-(3\er15)$^\circ$ \\
\multicolumn{2}{l}{Residue $\pi N\to \Delta\eta$ \hfill  18\er5} &
Phase & -(60\er15)$^\circ$
\\[0.3ex]\hline\\[-2ex]
\multicolumn{2}{l}{$A^{1/2}$ (\,GeV$^{-\frac12}$)\hfill 0.170\er0.020}& Phase &(50\er15)$^\circ$\\
\multicolumn{2}{l}{$A^{3/2}$ (\,GeV$^{-\frac12}$)\hfill 0.170\er0.025}& Phase &(45\er10)$^\circ$\\
\hline\hline\\[-1.5ex]
\multicolumn{4}{l}{$\Delta(1700)\frac32^-$ Breit-Wigner parameters (MeV) }\\[0.3ex]\hline\\[-2ex]
$M_{\rm BW}$& $1715^{+30}_{-15}$ & $\Gamma_{\rm BW}$& $310^{+40}_{-15}$\\
Br($N\pi $)   &  22\er4\% &
Br($\Delta\eta$) &  5\er2\%\\
Br($\Delta\pi_{L\!=\!0})$ &  22\er14\%& Br($\Delta\pi_{L\!=\!2})$ &
12\er10\%
\\[0.3ex]\hline\\[-2ex]
\multicolumn{2}{l}{$A_{BW}^{1/2}$
(\,GeV$^{-\frac12}$)\hfill~0.160\er0.020}&
\multicolumn{2}{l}{$A_{BW}^{3/2}$ (\,GeV$^{-\frac12}$)\hfill~0.165\er0.025} \\
\hline\hline\\
 \multicolumn{4}{l}{\boldmath
\fbox{\fbox{$\Delta(1905)\frac52^+$}}\unboldmath\qquad
or \quad $\Delta(1905)F_{35}$}  \\[-0.7ex]
&&&\\\hline\hline\\[-1.5ex]
\multicolumn{4}{l}{$\Delta(1905)\frac52^+$ pole parameters (MeV) }\\[0.3ex]\hline\\[-2ex]
$M_{\rm pole}$ &\hfill 1805\er10& $\Gamma_{\rm pole}$&\hfill 300\er15\\
\multicolumn{2}{l}{Elastic pole residue \hfill 20\er2}& Phase &-(44\er5)$^\circ$ \\
\multicolumn{2}{l}{Residue $\pi N\to \Delta\pi_{L\!=\!1}$ \hfill
37\er7} & Phase & (0\er15)$^\circ$
\\[0.3ex]\hline\\[-2ex]
\multicolumn{2}{l}{$A^{1/2}$ (\,GeV$^{-\frac12}$)\hfill~0.025\er0.005}& Phase &-(23\er15)$^\circ$\\
\multicolumn{2}{l}{$A^{3/2}$ (\,GeV$^{-\frac12}$)\hfill-0.050\er0.004}& Phase &(0\er10)$^\circ$\\
\hline\hline\\[-1.5ex]
\multicolumn{4}{l}{$\Delta(1905)\frac52^+$ Breit-Wigner parameters (MeV) }\\[0.3ex]\hline\\[-2ex]
$M_{\rm BW}$& 1861\er6 & $\Gamma_{\rm BW}$& 335\er18\\
Br($N\pi $)   &  13\er2\% &
Br($\Delta\pi_{L\!=\!1})$ &  45\er10\%\\
[0.3ex]\hline\\[-2ex]
\multicolumn{2}{l}{$A_{BW}^{1/2}$
(\,GeV$^{-\frac12}$)\hfill~0.025\er0.005}\\
\multicolumn{2}{l}{$A_{BW}^{3/2}$ (\,GeV$^{-\frac12}$)\hfill~-0.049\er0.004} \\
\hline\hline\\
\multicolumn{4}{l}{\boldmath
\fbox{\fbox{$\Delta(1920)\frac32^+$}}\unboldmath\qquad
or \quad $\Delta(1920)P_{33}$}  \\[-0.7ex]
&&&\\\hline\hline\\[-1.5ex]
\multicolumn{4}{l}{$\Delta(1920)\frac32^+$ pole parameters (MeV) }\\[0.3ex]\hline\\[-2ex]
$M_{\rm pole}$ &\hfill 1890\er30& $\Gamma_{\rm pole}$&\hfill 300\er60\\
\multicolumn{2}{l}{Elastic pole residue \hfill 17\er8}& Phase &-(40\er20)$^\circ$ \\
\multicolumn{2}{l}{Residue $\pi N\to \Sigma K$ \hfill  14\er7} & Phase &  (80\er40)$^\circ$\\
\multicolumn{2}{l}{Residue $\pi N\to \Delta\eta$ \hfill  27\er12} & Phase &  (70\er20)$^\circ$ \\
\multicolumn{2}{l}{Residue $\pi N\to \Delta\pi_{L\!=\!1}$ \hfill 30\er13} & Phase & -(120\er30)$^\circ$\\
\multicolumn{2}{l}{Residue $\pi N\to \Delta\pi_{L\!=\!3}$ \hfill
44\er14} & Phase & -(95\er35)$^\circ$
\\[0.3ex]\hline\\[-2ex]
\multicolumn{2}{l}{$A^{1/2}$ (\,GeV$^{-\frac12}$)\hfill $0.130^{+0.030}_{-0.060}$}& Phase &-(65\er20)$^\circ$\\
\multicolumn{2}{l}{$A^{3/2}$ (\,GeV$^{-\frac12}$)\hfill $0.115^{+0.025}_{-0.050}$}& Phase &-(160\er20)$^\circ$\\
\hline\hline\\[-1.5ex]
\multicolumn{4}{l}{$\Delta(1920)\frac32^+$ Breit-Wigner parameters (MeV) }\\[0.3ex]\hline\\[-2ex]
$M_{\rm BW}$& 1900\er30 & $\Gamma_{\rm BW}$& 310\er60\\
Br($N\pi $)   &  8\er4\% &Br($\Sigma K$)   &  4\er2\%\\
Br($\Delta\eta$) &  15\er8\%\\
Br($\Delta\pi_{L\!=\!1}$) & 22\er9\% &Br($\Delta\pi_{L\!=\!3}$)    & 45\er14\%\\
[0.3ex]\hline\\[-2ex]
\multicolumn{2}{l}{$A_{BW}^{1/2}$
(\,GeV$^{-\frac12}$)\hfill~$0.130^{+0.030}_{-0.060}$}\\
\multicolumn{2}{l}{$A_{BW}^{3/2}$ (\,GeV$^{-\frac12}$)\hfill~-$0.115^{+0.025}_{-0.050}$} \\
\hline\hline\\
\hspace{-2mm}
\end{tabular}
\end{tabular}
\end{table*}
\clearpage
\begin{table}[pt]
\hspace{-2mm}\begin{tabular}{lrlr} \multicolumn{4}{l}{\boldmath
\fbox{\fbox{$\Delta(1940)\frac32^-$}}\unboldmath\qquad
or \quad $\Delta(1940)D_{33}$}  \\[-0.7ex]
&&&\\\hline\hline\\[-1.5ex]
\multicolumn{4}{l}{$\Delta(1940)\frac32^-$ pole parameters (MeV) }\\[0.3ex]\hline\\[-2ex]
$M_{\rm pole}$ &\hspace{13mm}$1990^{+100}_{-\ 50}$& $\Gamma_{\rm pole}$&\hspace{16mm}450\er90\\
\multicolumn{2}{l}{Elastic pole residue \hfill 4\er4}& Phase & \\
\\[0.3ex]\hline\\[-2ex]
~\\
~\\
[0.8ex]
\hline\hline\\[-1.5ex]
\multicolumn{4}{l}{$\Delta(1940)\frac32^-$ Breit-Wigner parameters (MeV) }\\[0.3ex]\hline\\[-2ex]
$M_{\rm BW}$& $1995^{+105}_{-\ 60}$ & $\Gamma_{\rm BW}$& 450\er100\\
~\\
~\\
[0.3ex]\hline\\[-2ex]
~\\
~\\
[0.8ex] \hline\hline
\end{tabular}
\end{table}

 \noindent deed, we find a strong $N(1720)3/2^+\to
N(1520)3/2^-\pi$ coupling. There seems to be a sizable
$N(1720)3/2^+\to \Lambda K$ coupling as well; the latter decay
requires L$=1$. $N(1710)1/2^+$ may also have a significant $\Lambda
K$ coupling. A detailed study is required of the analytic structure
of these two resonances in the threshold region. We have not
included $\Delta(1750)1/2^+$ in the Tables below. We find no trace
of evidence for this resonance and doubt that it exists. At present,
the results on $\Delta(1940)3/2^-$ from $\gamma p\to p2\pi^0$ and
$\gamma p\to p\pi^0\eta$ are not consistent. Also this issue needs
further studies. At present, we give generous errors.

A few ``new" resonances are reported. ``New"  does not mean, that
resonances with these quantum numbers and similar masses and widths
have not been reported before. But so far, these resonances have not
been included in the Review of Particle Properties. These resonances
are\bc
$N(1880)\frac12^+$, $N(1860)\frac52^+$,$ N(1895)\frac12^-$, \\
$N(1875)\frac32^-$, $N(2150)\frac32^-$, and $N(2060)\frac52^-$.\ec

Yet, $N(2150)3/2^-$ could be the 2* resonance $N(2080)3/2^-$, and
$N(2060)5/2^-$ could be related to $N(2200)5/2^-$, with 2* as well,
of the Particle Data Group.

The $N(1880)1/2^+$ resonance was first suggested when data on
$\gamma p\to \Sigma^+K^0_s$ from the CBELSA collaboration
\cite{Castelijns:2007qt} were included in the BnGa partial wave
analysis. $N(1975)3/2^+$ emer\-ges from BNGA2011-02 only; it was
first reported in \cite{Anisovich:2011ye}. Early evidence for
$N(1860)5/2^+$ has been reported with Breit-Wigner parameters
$(M_{\rm BW}; \Gamma_{\rm BW})$ equal to $(1882\pm 10; 95\pm 20)$
\cite{Hohler:1979yr,Arndt:2006bf}, $1903\pm 87; 490\pm 310)$
\cite{Manley:1992yb}, and $(1817.7; 117.6)$ \cite{Arndt:2006bf}.
Evidence for  $N(1895)1/2^-$ has been reported by H\"ohler {\it et
al.} \cite{Hohler:1984ux} giving Breit-Wigner parameters of $M_{\rm
BW}= 1880\pm20, \Gamma_{\rm BW} =95\pm 30$\,MeV for a pole in the
$I(J^P)=1/2(1/2^-)$ wave. Manley {\it et al.} \cite{Manley:1992yb}
found a broad state, $M_{\rm BW}= 1928\pm59, \Gamma_{\rm BW} =414\pm
157$\,MeV. Vrana {\it et al.} \cite{Vrana:1999nt} reported $M_{\rm
BW}= 1822\pm43, \Gamma_{\rm BW} =246\pm185$\,MeV. A third and a
forth pole in the $I(J^P)=1/2(1/2^-)$ wave was suggested in
\cite{Tiator:2010rp}. The third pole was given with mass and width
of $M_{\rm pole}=1733$\,MeV; $\Gamma_{\rm pole}= 180$\,MeV, and in
\cite{Hadzimehmedovic:2011ua} with $M_{\rm pole}=1745\pm80$;
$\Gamma_{\rm pole}= 220\pm95$\,MeV. The latter pole was also seen by
Cutkosky {\it et al.} \cite{Cutkosky:1980rh} at $M_{\rm pole}=
2150\pm70, \Gamma_{\rm pole} =350\pm 100$\,MeV and confirmed by
Tiator {\it et al.} \cite{Tiator:2010rp}.
\begin{table}[pt]
\hspace{-2mm}\begin{tabular}{lrlr} \multicolumn{4}{l}{\boldmath
\fbox{\fbox{$\Delta(1950)\frac72^+$}}\unboldmath\qquad
or \quad $\Delta(1950)F_{37}$}  \\[-0.7ex]
&&&\\\hline\hline\\[-1.5ex]
\multicolumn{4}{l}{$\Delta(1950)\frac72^+$ pole parameters (MeV) }\\[0.3ex]\hline\\[-2ex]
$M_{\rm pole}$ &\hspace{12mm}1890\er4& $\Gamma_{\rm pole}$&\hspace{9mm}243\er8\\
\multicolumn{2}{l}{Elastic pole residue \hfill 58\er2}& Phase &-(24\er3)$^\circ$ \\
\multicolumn{2}{l}{Residue $\pi N\to \Sigma K$ \hfill  6\er1} & Phase &-(65\er25)$^\circ$ \\
\multicolumn{2}{l}{Residue $\pi N\to \Delta\pi_{L\!=\!3}$ \hfill
14\er4} & Phase &(12\er10)$^\circ$
\\[0.3ex]\hline\\[-2ex]
\multicolumn{2}{l}{$A^{1/2}$ (\,GeV$^{-\frac12}$)\hfill -0.072\er0.004}& Phase &-(7\er5)$^\circ$\\
\multicolumn{2}{l}{$A^{3/2}$ (\,GeV$^{-\frac12}$)\hfill -0.096\er0.005}& Phase &-(7\er5)$^\circ$\\
\hline\hline\\[-1.5ex]
\multicolumn{4}{l}{$\Delta(1950)\frac72^+$ Breit-Wigner parameters
(MeV)
}\\[0.3ex]\hline\\[-2ex]
$M_{\rm BW}$& 1915\er6 & $\Gamma_{\rm BW}$&246\er10 \\
Br($N\pi $)   &  45\er2\% &Br($\Sigma K $)   &  0.4\er0.1\%\\
Br($\Delta\pi_{L\!=\!3})$ &  2.8\er1.4\% \\
[0.3ex]\hline\\[-2ex]
\multicolumn{2}{l}{$A_{BW}^{1/2}$ (\,GeV$^{-\frac12}$)\hfill
~-0.071\er0.004}&\multicolumn{2}{l}{$A_{BW}^{3/2}$
(\,GeV$^{-\frac12}$)\hfill
~-0.094\er0.005}\\
\hline\hline
\end{tabular}
\end{table}

In the $\frac12(\frac32^-)$ wave, Cutkosky {\it et al.}
\cite{Cutkosky:1980rh} reported two resonances, the lower mass state
at  $M_{\rm BW}= 1880\pm100, \Gamma_{\rm BW} =180\pm 60$\,MeV, the
higher mass pole at $M_{\rm BW}= 2060\pm60, \Gamma_{\rm BW} =300\pm
1ß0$\,MeV. Saxon {\it et al.} \cite{RL-79-055} and Bell {\it et al.}
\cite{RL-83-005} observed a $\frac12(\frac32^-)$ resonance in the
reaction $\pi^-p\to \Lambda K^0$ at $(1900; 240)$\,MeV and $(1920;
320)$\,MeV, respectively. Based on SAPHIR data on $\gamma p\to
\Lambda K^+$ \cite{Glander:2003jw}, Mart and Bennhold claimed
evidence for a $\frac12(\frac32^-)$ resonance at 1895\,MeV
\cite{Mart:1999ed} which was confirmed by us on a richer data base
in \cite{Anisovich:2005tf,Sarantsev:2005tg}, with mass and width of
$(1875\pm25; 80\pm 20)$\,MeV, respectively. The high-mass
$N_{3/2^-}$ was also seen in
\cite{Anisovich:2005tf,Sarantsev:2005tg} with $(2166^{+25}_{-50};
\Gamma=300\pm65)$\,MeV and in \cite{Schumacher:2010qx} with
$(2100\pm20; 200\pm50)$\,MeV.

\section{Significance and rating}
In  Table \ref{sign} we give our rating of the evidence with which
baryon resonances are observed. By definition,\\[0.5ex]
\begin{tabular}{ll}
 ****& Existence is certain, and properties are at least fairly\\
     &  well explored.\\
 ***& Existence ranges from very likely to certain, but fur-\\
    & ther confirmation is desirable and/or quantum num- \\
    & bers, branching fractions {\it etc.}  are not well determined.\\
 **& Evidence of existence is only fair.\\
 *&Evidence of existence is poor.\\[0.5ex]
\end{tabular}
The significance of a resonance and of its decay modes is estimated
from three sources: (i) from the increase in $\chi^2$ when a
resonance is removed from the fit, both the overall increase in
$\chi^2$, and the increase in $\chi^2$ in specific final states,
(ii) from the stability of the fit result when the hypothesis (e.g.
number of poles in a given partial wave) is changed, and (iii) from
the errors in the definition of masses, widths, residues,
photocouplings, {\it etc.} As a rule we give 1* when a decay mode is
seen with a significance of $2\sigma$, 2* for a significance of
$3.5\sigma$, and 3* for a significance of $5\sigma$. As there are
ambiguous solutions, we do not assign 4* for decays derived from
photoproduction. In some cases, the errors are large, and the
significance is high.
\begin{table}[pt]
\caption{\label{sign}Star rating suggested for baryon resonances and
their decays. Ratings of the Particle Data Group are given as \ *;
additional stars suggested from this analysis are represented by \
$\star$; (*)  stands for stars which should be removed. }
\renewcommand{\arraystretch}{1.15}
\begin{tabular}{lcccccccc}
\hline\hline
 \hspace{-3mm}&\hspace{-3mm}    all     \hspace{-3mm}&\hspace{-3mm}$\pi N$\hspace{-3mm}&\hspace{-3mm}$\gamma N$\hspace{-3mm}&\hspace{-3mm} $N\eta$ \hspace{-3mm}&\hspace{-3mm}$\Lambda K$\hspace{-3mm}&\hspace{-3mm}$\Sigma K$ \hspace{-3mm}&\hspace{-3mm}$\Delta\pi$ \hspace{-3mm}&\hspace{-3mm}$N\sigma$\\
\hline
$N(1440)\frac12^+$\hspace{-3mm}&\hspace{-3mm}****        \hspace{-3mm}&\hspace{-3mm}****   \hspace{-3mm}&\hspace{-3mm}***\rs\hspace{-3mm}&\hspace{-3mm}\rmstar\hspace{-3mm}&\hspace{-3mm}\hspace{-3mm}&\hspace{-3mm}\hspace{-3mm}&\hspace{-3mm}***\hspace{-3mm}&\hspace{-3mm}\threestar  \\
$N(1710)\frac12^+$\hspace{-3mm}&\hspace{-3mm}***         \hspace{-3mm}&\hspace{-3mm}***    \hspace{-3mm}&\hspace{-3mm}***   \hspace{-3mm}&\hspace{-3mm}**\rs\hspace{-3mm}&\hspace{-3mm}\threestar\hspace{-3mm}&\hspace{-3mm}\twostar\hspace{-3mm}&\hspace{-3mm}*\rmstar\hspace{-3mm}&\hspace{-3mm}  \\
$N(1880)\frac12^+$\hspace{-3mm}&\hspace{-3mm}\twostar   \hspace{-3mm}&\hspace{-3mm}\rs\hspace{-3mm}&\hspace{-3mm}\rs\hspace{-3mm}&\hspace{-3mm} \hspace{-3mm}&\hspace{-3mm}\twostar\hspace{-3mm}&\hspace{-3mm}$\rs$\hspace{-3mm}&\hspace{-3mm}\hspace{-3mm}&\hspace{-3mm}     \\
\hline
$N(1535)\frac12^-$\hspace{-3mm}&\hspace{-3mm}****        \hspace{-3mm}&\hspace{-3mm}****   \hspace{-3mm}&\hspace{-3mm}****  \hspace{-3mm}&\hspace{-3mm}****\hspace{-3mm}&\hspace{-3mm} \hspace{-3mm}&\hspace{-3mm}\hspace{-3mm}&\hspace{-3mm}*\hspace{-3mm}&\hspace{-3mm}    \\
$N(1650)\frac12^-$\hspace{-3mm}&\hspace{-3mm}****        \hspace{-3mm}&\hspace{-3mm}****   \hspace{-3mm}&\hspace{-3mm}***   \hspace{-3mm}&\hspace{-3mm}*\rts \hspace{-3mm}&\hspace{-3mm}*** \hspace{-3mm}&\hspace{-3mm}** \hspace{-3mm}&\hspace{-3mm} **\rmstar \hspace{-3mm}&\hspace{-3mm}    \\
$N(1895)\frac12^-$\hspace{-3mm}&\hspace{-3mm}\twostar    \hspace{-3mm}&\hspace{-3mm}$\rs$\hspace{-3mm}&\hspace{-3mm}\twostar \hspace{-3mm}&\hspace{-3mm}$\twostar$\hspace{-3mm}&\hspace{-3mm}\twostar \hspace{-3mm}&\hspace{-3mm}\rs      \\
\hline
$N(1720)\frac32^+$\hspace{-3mm}&\hspace{-3mm}****        \hspace{-3mm}&\hspace{-3mm}****   \hspace{-3mm}&\hspace{-3mm}****  \hspace{-3mm}&\hspace{-3mm}****\hspace{-3mm}&\hspace{-3mm}**\hspace{-3mm}&\hspace{-3mm}**\hspace{-3mm}&\hspace{-3mm}***\hspace{-3mm}&\hspace{-3mm}    \\
$N(1900)\frac32^+$\hspace{-3mm}&\hspace{-3mm}**\rs       \hspace{-3mm}&\hspace{-3mm}**\hspace{-3mm}&\hspace{-3mm}\threestar\hspace{-3mm}&\hspace{-3mm}\twostar\hspace{-3mm}&\hspace{-3mm}\threestar\hspace{-3mm}&\hspace{-3mm}\twostar \hspace{-3mm}&\hspace{-3mm}\twostar\hspace{-3mm}&\hspace{-3mm}       \\
\hline
$N(1520)\frac32^-$\hspace{-3mm}&\hspace{-3mm}****        \hspace{-3mm}&\hspace{-3mm}****   \hspace{-3mm}&\hspace{-3mm}****  \hspace{-3mm}&\hspace{-3mm}***\hspace{-3mm}&\hspace{-3mm} \hspace{-3mm}&\hspace{-3mm}\hspace{-3mm}&\hspace{-3mm}****\hspace{-3mm}&\hspace{-3mm}    \\
$N(1700)\frac32^-$\hspace{-3mm}&\hspace{-3mm}**\rs    \hspace{-3mm}&\hspace{-3mm}**\hspace{-3mm}&\hspace{-3mm}**    \hspace{-3mm}&\hspace{-3mm}*\hspace{-3mm}&\hspace{-3mm}*\rmstar\hspace{-3mm}&\hspace{-3mm}* \hspace{-3mm}&\hspace{-3mm}**\rs\hspace{-3mm}&\hspace{-3mm}       \\
$N(1875)\frac32^-$\hspace{-3mm}&\hspace{-3mm}\threestar  \hspace{-3mm}&\hspace{-3mm}$\rs$\hspace{-3mm}&\hspace{-3mm}\threestar\hspace{-3mm}&\hspace{-3mm}\hspace{-3mm}&\hspace{-3mm}\threestar \hspace{-3mm}&\hspace{-3mm}\twostar  \hspace{-3mm}&\hspace{-3mm}\hspace{-3mm}&\hspace{-3mm}\threestar   \\
$N(2150)\frac32^-$\hspace{-3mm}&\hspace{-3mm}\twostar    \hspace{-3mm}&\hspace{-3mm} $\twostar$ \hspace{-3mm}&\hspace{-3mm}\twostar      \hspace{-3mm}&\hspace{-3mm}\hspace{-3mm}&\hspace{-3mm}\twostar  \hspace{-3mm}&\hspace{-3mm}\hspace{-3mm}&\hspace{-3mm} \twostar  \hspace{-3mm}&\hspace{-3mm}  \\
\hline
$N(1680)\frac52^+$\hspace{-3mm}&\hspace{-3mm}****        \hspace{-3mm}&\hspace{-3mm}****   \hspace{-3mm}&\hspace{-3mm}****  \hspace{-3mm}&\hspace{-3mm} *\hspace{-3mm}&\hspace{-3mm}\hspace{-3mm}&\hspace{-3mm}\hspace{-3mm}&\hspace{-3mm}**\rmstar\hspace{-3mm}&\hspace{-3mm} $\rs\rs$\\
$N(1860)\frac52^+$\hspace{-3mm}&\hspace{-3mm}$\rs$     \hspace{-3mm}&\hspace{-3mm}$\rs$\hspace{-3mm}&\hspace{-3mm}$\rs$\hspace{-3mm}&\hspace{-3mm}   \\
$N(2000)\frac52^+$\hspace{-3mm}&\hspace{-3mm}*\twostar  \hspace{-3mm}&\hspace{-3mm}*\rmstar \hspace{-3mm}&\hspace{-3mm}$\twostar$  \hspace{-3mm}&\hspace{-3mm}\twostar \hspace{-3mm}&\hspace{-3mm}\twostar\hspace{-3mm}&\hspace{-3mm}\rs\hspace{-3mm}&\hspace{-3mm}\hspace{-3mm}&\hspace{-3mm}  \\
\hline
$N(1675)\frac52^-$\hspace{-3mm}&\hspace{-3mm}****        \hspace{-3mm}&\hspace{-3mm}****   \hspace{-3mm}&\hspace{-3mm}***\rmstar  \hspace{-3mm}&\hspace{-3mm}*\hspace{-3mm}&\hspace{-3mm}*\hspace{-3mm}&\hspace{-3mm}\hspace{-3mm}&\hspace{-3mm}***\rmstar \hspace{-3mm}&\hspace{-3mm}$\rs$      \\
$N(2060)\frac52^-$\hspace{-3mm}&\hspace{-3mm}\threestar  \hspace{-3mm}&\hspace{-3mm}\twostar\hspace{-3mm}&\hspace{-3mm}\threestar\hspace{-3mm}&\hspace{-3mm}$\rs$\hspace{-3mm}&\hspace{-3mm}\hspace{-3mm}&\hspace{-3mm}\twostar\hspace{-3mm}&\hspace{-3mm}        \\
\hline
$N(1990)\frac72^+$\hspace{-3mm}&\hspace{-3mm}**        \hspace{-3mm}&\hspace{-3mm}*\rmstar   \hspace{-3mm}&\hspace{-3mm}\rs\rs \hspace{-3mm}&\hspace{-3mm} \\
\hline
$N(2190)\frac72^-$\hspace{-3mm}&\hspace{-3mm}****        \hspace{-3mm}&\hspace{-3mm}****   \hspace{-3mm}&\hspace{-3mm}*\rs\rs      \hspace{-3mm}&\hspace{-3mm} \hspace{-3mm}&\hspace{-3mm}\rs\rs\hspace{-3mm}&\hspace{-3mm}\hspace{-3mm}&\hspace{-3mm}\\
\hline
$N(2220)\frac92^+$\hspace{-3mm}&\hspace{-3mm}****        \hspace{-3mm}&\hspace{-3mm}****   \hspace{-3mm}&\hspace{-3mm}      \hspace{-3mm}&\hspace{-3mm} \\
\hline
$N(2250)\frac92^-$\hspace{-3mm}&\hspace{-3mm}****        \hspace{-3mm}&\hspace{-3mm}****   \hspace{-3mm}&\hspace{-3mm}      \hspace{-3mm}&\hspace{-3mm} \\
\hline
$\Delta(1910)\frac12^+$\hspace{-3mm}&\hspace{-3mm}****   \hspace{-3mm}&\hspace{-3mm}****   \hspace{-3mm}&\hspace{-3mm}*\rs\hspace{-3mm}&\hspace{-3mm} \hspace{-3mm}&\hspace{-3mm}\hspace{-3mm}&\hspace{-3mm}*\rs\hspace{-3mm}&\hspace{-3mm}*\rs   \\
\hline
$\Delta(1620)\frac12^-$\hspace{-3mm}&\hspace{-3mm}****   \hspace{-3mm}&\hspace{-3mm}****   \hspace{-3mm}&\hspace{-3mm}***   \hspace{-3mm}&\hspace{-3mm} \hspace{-3mm}&\hspace{-3mm}\hspace{-3mm}&\hspace{-3mm}\hspace{-3mm}&\hspace{-3mm}****\hspace{-3mm}&\hspace{-3mm}    \\
$\Delta(1900)\frac12^-$\hspace{-3mm}&\hspace{-3mm}**     \hspace{-3mm}&\hspace{-3mm}**     \hspace{-3mm}&\hspace{-3mm}*\rs\hspace{-3mm}&\hspace{-3mm} \hspace{-3mm}&\hspace{-3mm}\hspace{-3mm}&\hspace{-3mm} *\rs\hspace{-3mm}&\hspace{-3mm}*\rs\hspace{-3mm}&\hspace{-3mm} \\
\hline
$\Delta(1232)\frac32^+$\hspace{-3mm}&\hspace{-3mm}****   \hspace{-3mm}&\hspace{-3mm}****   \hspace{-3mm}&\hspace{-3mm}****  \hspace{-3mm}&\hspace{-3mm}  \\
$\Delta(1600)\frac32^+$\hspace{-3mm}&\hspace{-3mm}***    \hspace{-3mm}&\hspace{-3mm}***    \hspace{-3mm}&\hspace{-3mm}**$\rs$\hspace{-3mm}&\hspace{-3mm} \hspace{-3mm}&\hspace{-3mm}\hspace{-3mm}&\hspace{-3mm} &***\hspace{-3mm}&\hspace{-3mm}  \\
$\Delta(1920)\frac32^+$\hspace{-3mm}&\hspace{-3mm}***    \hspace{-3mm}&\hspace{-3mm}***    \hspace{-3mm}&\hspace{-3mm}*\rs\hspace{-3mm}&\hspace{-3mm} \hspace{-3mm}&\hspace{-3mm} \hspace{-3mm}&\hspace{-3mm}*\rs\rs\hspace{-3mm}&\hspace{-3mm}**   \\
\hline
$\Delta(1700)\frac32^-$\hspace{-3mm}&\hspace{-3mm}***    \hspace{-3mm}&\hspace{-3mm}***    \hspace{-3mm}&\hspace{-3mm}***\hspace{-3mm}&\hspace{-3mm} \hspace{-3mm}&\hspace{-3mm}\hspace{-3mm}&\hspace{-3mm}\hspace{-3mm}&\hspace{-3mm}**\hspace{-3mm}&\hspace{-3mm}   \\
$\Delta(1940)\frac32^-$\hspace{-3mm}&\hspace{-3mm}*    \hspace{-3mm}&\hspace{-3mm}*      \hspace{-3mm}&\hspace{-3mm}\rs\rs\hspace{-3mm}&\hspace{-3mm} \hspace{-3mm}&\hspace{-3mm}\hspace{-3mm}&\hspace{-3mm}\hspace{-3mm}&\multicolumn{2}{c}{$\vert$\rs\ from $\Delta\eta$$\vert$}  \\
\hline
$\Delta(1905)\frac52^+$\hspace{-3mm}&\hspace{-3mm}****   \hspace{-3mm}&\hspace{-3mm}****   \hspace{-3mm}&\hspace{-3mm}****  \hspace{-3mm}&\hspace{-3mm}\hspace{-3mm}&\hspace{-3mm}\hspace{-3mm}&\hspace{-3mm}*\rs\rs\hspace{-3mm}&\hspace{-3mm}**\rmtstar\\
\hline
$\Delta(1950)\frac72^+$\hspace{-3mm}&\hspace{-3mm}****   \hspace{-3mm}&\hspace{-3mm}****   \hspace{-3mm}&\hspace{-3mm}***   \hspace{-3mm}&\hspace{-3mm}\hspace{-3mm}&\hspace{-3mm}\hspace{-3mm}&\hspace{-3mm}*\rs\rs\hspace{-3mm}&\hspace{-3mm}**\rs\\
\hline \hline
\end{tabular}
\renewcommand{\arraystretch}{1.0}
\end{table}
This happens, if there are two solutions which give different values
for an observable, e.g. for its photoproduction amplitude. Without
the resonance, the photoproduction data cannot be described; hence
we are sure that the resonance is needed. But the actual value may
be less certain. The star rating reflects our estimate how safe we
are in claiming the existence of the resonance from photoproduction
data; the error gives the range of values of resonance properties
which might be assigned to a given resonance.

\subsection*{Acknowledgements}
We would like to thank the members of SFB/TR16 for continuous
encouragement. We acknowledge support from the Deutsche
Forschungsgemeinschaft (DFG) within the SFB/ TR16 and from the
Forschungszentrum J\"ulich within the FFE program.

\end{document}